\documentclass[11pt]{article}
\usepackage{jhip}

\newcommand{\nc}{\newcommand}
\nc{\rnc}{\renewcommand }
\rnc{\d}{\mathrm{d}}
\nc{\D}{\partial}
\nc{\g}{\gamma}
\nc{\n}{{(n)}}
\rnc{\t}{\tau}
\nc{\ep}{\epsilon}
\nc{\K}{\mathcal{K}}
\nc{\lrarrow}{\leftrightarrow}
\nc{\nn}{\nonumber}
\renewcommand{\[}{\begin{equation}}
\renewcommand{\]}{\end{equation}}
\newcommand{\bea}{\begin{eqnarray}}
\newcommand{\eea}{\end{eqnarray}}
\renewcommand{\(}{\left(}
\renewcommand{\)}{\right)}

\newcommand{\p}{\partial}
\newcommand{\s}{\sigma}
\newcommand{\eps}{\epsilon}

\title{The holographic fluid dual to vacuum Einstein gravity}

\author[a]{Geoffrey Comp\`ere,}
\author[b]{Paul McFadden,}
\author[a,b,c]{Kostas Skenderis}
\author[b,c]{and Marika Taylor}

\affiliation[a]{KdV Institute for Mathematics,}
\affiliation[b]{Institute for Theoretical Physics,}
\affiliation[c]{Gravitation and Astro-Particle Physics Amsterdam,\\[0.5ex]
Science Park 904, 1090 GL Amsterdam, the Netherlands.}

\emailAdd{gcompere@uva.nl}
\emailAdd{P.L.McFadden@uva.nl}
\emailAdd{K.Skenderis@uva.nl}
\emailAdd{M.Taylor@uva.nl}

\abstract{

We present an algorithm for systematically reconstructing a solution of the $(d+2)$-dimensional vacuum Einstein equations
from a $(d+1)$-dimensional fluid, extending the non-relativistic hydrodynamic
expansion of Bredberg et al.~in arXiv:1101.2451 to arbitrary order. The fluid satisfies equations of motion
which are the incompressible Navier-Stokes equations, corrected by specific higher-derivative terms. The uniqueness and regularity of this solution
is established to all orders and explicit results are given for the bulk metric and the stress tensor of the dual fluid through fifth order in the
hydrodynamic expansion. We establish the validity of a relativistic hydrodynamic description for the dual fluid, which has the unusual property
of having a vanishing equilibrium energy density. The gravitational results are used to identify transport coefficients of the dual fluid, which
also obeys an interesting and exact constraint on its stress tensor. We
propose novel Lagrangian models which realise key properties of the holographic fluid.

}

\begin{document}
\maketitle


\section{Introduction and summary of results}

Black hole physics suggests that gravitational theories are holographic
and indeed over the last 15 years we have witnessed remarkable progress in
uncovering holographic relations, mainly between $(d+2)$-dimensional gravitational theories
with negative cosmological constant and $(d+1)$-dimensional local quantum
field theories. The case of asymptotically flat spacetimes remains elusive
and indeed the structure of the asymptotic solutions
\cite{BS,BS2,deHaro:2000wj} and the divergences of the
on-shell action  suggest \cite{Skenderis:2002wp} that
if there is a dual theory it is nonlocal, see also \cite{Papadimitriou:2010as}.

In a recent paper \cite{Bredberg} (see
\cite{Andy, Fouxon:2008tb,Bhattacharyya:2008kq,Eling:2009pb, Padmanabhan, Damour1, Damour2, Thorne} for earlier
relevant work),
a remarkable relation between
incompressible non-relativistic fluids in $(d+1)$ dimensions
satisfying the Navier-Stokes equation and
$(d+2)$-dimensional Ricci
flat metrics was found. The construction of \cite{Bredberg} starts
from considering the portion of Minkowski spacetime between a flat
hypersurface $\Sigma_c$, given by the equation $X^2 - T^2 = 4 r_c$, and
its future horizon ${\cal H}^+$, the null surface $X=T$. Then the effects of finite perturbations
of the extrinsic curvature
of $\Sigma_c$, with the intrinsic metric $\g_{ab}$ kept fixed, were studied and it was found that
a regular Ricci flat metric exists provided that the Brown-York
stress tensor on $\Sigma_c$ is that of an incompressible Navier-Stokes
fluid. More precisely, they work in a hydrodynamic non-relativistic
limit and construct the bulk metric up to third order in the
hydrodynamic expansion. Apart from its intrinsic value, this
relationship provides a potential example of a holographic duality
involving a flat spacetime and as such it should be further explored.

The aim of this paper is two-fold. Firstly, we would like to provide
a systematic construction of the bulk solution to all orders in the hydrodynamic
expansion and secondly we would like to extract the physical properties of the
dual fluid in order to obtain clues about the nature of the dual theory.

Let us summarise our results. First recall that in the hydrodynamic limit
essentially any $(d+1)$-dimensional fluid is governed by
the incompressible Navier-Stokes equation
\[ \label{NS}
E_{i} \equiv \D_\t v_i + v^j\D_jv_i-\eta\D^2 v_i +\D_i P = 0, \qquad
E \equiv \D_i v^i = 0, \quad (i = 1, \ldots, d),
\]
where $v^i$ is the velocity field, $P$ is a pressure fluctuation about the
background value and $\eta$ is the kinematic viscosity.
This equation has a scaling symmetry and admits a one-parameter family of
solutions
\[
v_i^{(\ep)}(\t,\vec{x}) = \ep v_i(\ep^2\t,\ep\vec{x}), \qquad P^{(\ep)}(\t,\vec{x})=\ep^2 P(\ep^2\t,\ep \vec{x}).
\]
If one considers now higher-derivative corrections to $E_{i}$ and
$E$, these are naturally organised according to their scaling with
$\ep$.

Our first main result is an algorithmic construction of a
regular Ricci-flat metric corresponding to a fluid satisfying
(\ref{NS}) with specific higher-derivative corrections. We begin with the
observation that
the metric up to order $\ep^2$ with constant velocity and pressure fields
is actually flat space in disguise: it can be
obtained from the Rindler metric by a linear coordinate transformation
combining a boost which introduces a constant velocity field
$v^i$ with a simple shift and rescaling of the radial and time coordinates that
introduces a constant pressure function $P$. In fact, these transformations
are the most general infinitesimal diffeomorphisms that preserve the
induced metric at $r = r_c$ and lead to a Brown-York stress tensor which
is of the form of a fluid in equilibrium. (Recall that the Brown-York stress
tensor for Rindler spacetime is of that form.)

To extend the solution to next order in the hydrodynamic expansion,
we follow \cite{Bhattacharyya:2008jc} and promote
the velocity field and the pressure function to be
spacetime dependent quantities, subject to the requirement that the
Einstein equations hold up to order $\ep^3$.  To satisfy this
requirement one needs to introduce new terms of order $\ep^3$ in the
metric and we find that (\ref{NS})
holds. This reproduces the metric
given in \cite{Bredberg} (up to a choice of gauge for the dual fluid).

Next we prove inductively that
Einstein equations can in fact be satisfied to arbitrarily high order in $\ep$,
by adding appropriate terms to the metric and modifying the
Navier-Stokes equation and the incompressibility condition by specific higher
derivative corrections. In particular, the incompressibility
condition $E$ is modified at even powers of $\ep$ while the
Navier-Stokes equation $E_{i}$ is modified at odd powers of $\ep$.
We carried out this procedure explicitly to order $\ep^5$ and obtained
the resultant metric through to order $\ep^5$, given in
(\ref{g3})-(\ref{g41})-(\ref{g42})-(\ref{g5})
as well as the leading order corrections to $E$ and $E_{i}$, given in
(\ref{final_I}) and (\ref{final_NS}), respectively.

The dual fluid has a number of interesting and unconventional features.
Firstly, in equilibrium it has zero energy density and positive nonzero
pressure. In the hydrodynamic regime, the stress tensor receives
dissipative corrections,
the explicit form of which is given to order $\ep^5$ in
(\ref{T3}), (\ref{T4}) and (\ref{T5}). As a consequence of the (bulk)
Hamiltonian constraint, the fluid stress tensor satisfies
the quadratic constraint,
\[ \label{qua-con}
d T_{ab}T^{ab} = T^2.
\]
This is an interesting constraint on the dual theory which will be discussed
further below.

Since the bulk solution is constructed using a non-relativistic hydrodynamic
expansion in $\ep$, the fluid stress tensor appears to be
non-relativistic. In turns out however that the complete answer, including
dissipative terms, can be obtained from a relativistic hydrodynamic stress
tensor upon using the non-relativistic expansion in $\ep$. The rather
complicated expressions obtained by direct computation actually originate from
a much simpler relativistic stress tensor. The information
encoded in the stress tensor can then be expressed in terms of
a few transport coefficients. These considerations indicate also that
the underlying holographic dual theory should be relativistic.

Recall that a relativistic fluid stress tensor takes the general form
\[
 T_{ab} = \rho u_a u_b + p h_{ab} + \Pi^\perp_{ab}, \qquad u^a\Pi^\perp_{ab}=0,
\]
where $\rho$ and $p$ are the energy density and pressure of the fluid in
the local rest frame, $u^a$ is the relativistic fluid velocity (normalised such that $u_au^a=-1$),
and $h_{ab} = \g_{ab}+u_au_b$ is the induced metric on surfaces orthogonal
to the fluid velocity.  The term $\Pi^\perp_{ab}$ encodes dissipative
corrections and may be expanded in gradients of the fluid velocity.
When the equilibrium energy density is nonzero, one may show that
the Landau-Lifshitz prescription \cite{Landau} implies that the
energy density $\rho$ does not receive
any gradient corrections. This is in stark contrast to the case in hand
where the equilibrium fluid has zero energy density and the energy density becomes
nonzero and negative due to gradient terms. In fact the quadratic
constraint (\ref{qua-con}) uniquely determines $\rho$ in terms of the dissipative terms.
The leading order result for $\rho$ is given in (\ref{rho_lead}).

From the dissipative terms, $\Pi^\perp_{ab}$, one can read off the transport
coefficients. The relativistic corrections are organised according to
the number of derivatives they contain. Each of these terms can then be
expanded in the non-relativistic hydrodynamic expansion.
The $\ep$ expansion may increase the number of derivatives but it never
decreases them. This means that one can unambiguously extract all
transport coefficients but one will have to work to sufficiently high order
in $\ep$ in order to obtain all transport coefficients. At first order
in derivatives, there is only one possible term\footnote{For relativistic fluids in general
there would also be a bulk viscosity term.
For fluids with zero equilibrium energy density, this term vanishes
as a consequence of the equations of motion (see section \ref{Section:Hydro}).}
and from this the kinematic shear viscosity is obtained.
If we identify the entropy density of the fluid using the entropy of
the Rindler horizon, the ratio $\eta/s$ takes the universal value,
\begin{equation}
\eta/s = 1/(4 \pi).
\end{equation}
At second order, there are six transport coefficients (for a fluid in a flat background)
and our computation to order $\epsilon^5$ determines four of them, see
(\ref{PiT}) and (\ref{transport_coefficients}). The products of these coefficients with
the (background) pressure are pure numbers and one gets the following relations
\begin{equation}
2 c_1 p = c_2 p= c_3 p = c_4 p = -4.
\end{equation}
These results encode all the information contained in the dissipative parts of the
stress tensor, up to this order.

Let us now move to the non-dissipative part,
\begin{equation}\label{nondis}
T_{ab}^{\rm non{-}dissipative} = p h_{ab}.
\end{equation}
Note that this satisfies by itself the quadratic constraint.
One may ask whether there is a Lagrangian model that leads
to this stress tensor. It turns out that such a model indeed
exists and is given by\footnote{Interestingly, this
action has also been investigated
in the context of dark energy, see \cite{Niayesh1, Niayesh2}.}
\[
S = \int \d^{d+1}x \sqrt{-\g} \sqrt{-(\D\phi)^2}.
\]
This describes an ideal fluid
with  $u_a = \D_a \phi/\sqrt{-(\D\phi)^2}$ and stress tensor
given by (\ref{nondis}) with $p=\sqrt{-(\D\phi)^2}$. The equilibrium configuration
corresponds to the solution $\phi=\t$ and near-equilibrium configurations are obtained by looking at small
fluctuations around this solution. The background solution breaks the relativistic invariance and the expansion around this solution
leads to a  non-relativistic nonlocal action (since it contains an infinite number of derivatives).
Thus, this simple scalar theory (and its generalisations, see \eqref{gen-mod}) reproduces the non-dissipative
part of the stress tensor and incorporates many of the features ones expects
from the holographic dual theory.

\bigskip

The plan of the paper is as follows. In section \ref{eq-configurations}, we discuss a general class of flat metrics admitting a Rindler horizon
for which the Brown-York stress tensor on the constant-radius slice $\Sigma_c$ outside the horizon is that of a perfect fluid with zero energy density. These equilibrium
backgrounds are used in section \ref{Se:seed} as seed metrics to construct corresponding near-equilibrium configurations. In section \ref{Section_algorithm}, we
give the algorithm for systematically constructing the bulk solution to arbitrary order in the hydrodynamic expansion, while in section \ref{gauge-choice}, we explain
the gauge-fixing conditions imposed on the fluid. In section \ref{explicit}, we give explicit results for the metric and stress tensor up to order $\ep^5$, and
 in section \ref{Section:Hydro}, these results are interpreted in terms of transport coefficients for the dual fluid. Section \ref{Section:Models} proposes a simple
dual Lagrangian which captures key properties of the fluid, and in section \ref{Discussion}, we discuss our results. Finally, in the three appendices we present some of the technical results used in the main text.
In appendix \ref{app:radialgauge}, we discuss the choice of radial gauge;
in appendix \ref{app:no_rho}, we
derive the class of infinitesimal diffeomorphisms preserving the equilibrium form of the Brown-York stress tensor; and in appendix \ref{hydro_app}, we derive the general form
of the hydrodynamic expansion to second order in gradients for a relativistic fluid with vanishing equilibrium energy density.

\section{Equilibrium configurations}
\label{eq-configurations}

Let us first construct a class of Riemann-flat metrics which have the following properties: (i) they admit a co-dimension one hypersurface $\Sigma_c$ of flat induced metric
\[
\g_{ab}\d x^a\d x^b=-r_c\d\t^2+\d x_i\d x^i,\label{indmetric}
\]
where the speed of light $\sqrt{r_c}$ is arbitrary,
$x^\mu = (r, x^a)$ and $x^a=(\t,x^i)$ with $i=1,{\ldots},d$
(unless otherwise noted, Latin indices are raised with $\gamma^{ab}$ and Greek indices with $g^{\mu\nu}$);
(ii) the Brown-York stress tensor \cite{Brown:1992br} on $\Sigma_c$, given by\footnote{From here onwards we will set $16 \pi G = 1$.
We also assume the validity of classical gravity, i.e., higher-order curvature and quantum corrections are suppressed.}
\[
T_{ab}=\frac{1}{8\pi G}(K \gamma_{ab}-K_{ab}),
\]
where $K_{ab}$ is the extrinsic curvature of $\Sigma_c$, takes the form of a perfect fluid stress tensor,
\[
T_{ab} = \rho u_a u_b+ p h_{ab}\label{stresstensor},
\]
and finally; (iii) they are stationary with respect to $\p_\tau$ and homogeneous in the $x^i$ directions.

One such metric with these properties is
\[
\label{Rindler}
\d s^2 = \bar g_{\mu\nu}dx^\mu dx^\nu = -r\d\t^2+2\d\t\d r + \d x_i\d x^i,
\]
which describes flat space in ingoing Rindler coordinates. (Upon setting $\t=2\ln(X+T)$ and $4r=X^2-T^2$, we recover $\d s^2=-\d T^2+\d X^2$.)
The induced metric $\g_{ab}$ on the surface $\Sigma_c$ defined by $r=r_c$ then has the form \eqref{indmetric}, while the Brown-York stress tensor on $\Sigma_c$ has the form \eqref{stresstensor}, where $u_\tau = -r_c^{1/2}$, $u_i=0$, $\rho = 0$ and $p = r_c^{-1/2}$. The Rindler horizon is at $r=0$.

We may now obtain further metrics satisfying conditions (i), (ii) and (iii) by considering the action of diffeomorphisms on \eqref{Rindler}.
Since we wish to obtain a connected thermodynamical state space, it is sufficient to consider only diffeomorphisms that are continuously connected to the identity, i.e., those obtained by exponentiating infinitesimal diffeomorphisms.
In appendix \ref{app:no_rho}, we show that there are only two infinitesimal diffeomorphisms yielding metrics satisfying conditions (i), (ii) and (iii), after fixing the radial gauge as described in appendix \ref{app:radialgauge}.
Exponentiating these, we obtain the following two finite diffeomorphisms.

The first is a constant boost $\beta_i$, given by
\[
\sqrt{r_c}\t \rightarrow \g\sqrt{r_c}\t-\g\beta_i x^i, \qquad
x^i \rightarrow x^i -\g\beta^i\sqrt{r_c}\t+(\g-1)\frac{\beta^i\beta_j}{\beta^2}x^j,\label{t1}
\]
where $\g=(1-\beta^2)^{-1/2}$ and $\beta_i \equiv r_c^{-1/2} v_i$.  The second is a constant linear shift of $r$ and associated re-scaling of $\tau$,
\[
r \rightarrow r-r_h, \qquad \t \rightarrow (1-r_h/r_c)^{-1/2}\t.\label{t2}
\]
Note that this latter transformation shifts the position of the horizon to $r=r_h$; as we are only interested in the case where $\Sigma_c$ is outside the horizon we will restrict $r_h < r_c$.

Since these transformations are both linear in the coordinates, one can apply them in either order on Rindler spacetime. One obtains
\begin{align}
\d s^2 &= \frac{\d\tau^2}{1-v^2/r_c}\(v^2 - \frac{r-r_h}{1-r_h/r_c}\)+\frac{2\gamma}{\sqrt{1-r_h/r_c}}\d\tau \d r - \frac{2\gamma v_i}{r_c \sqrt{1-r_h/r_c}}\d x^i \d r \nn \\[1ex]
&\quad + \frac{2v_i}{1-v^2/r_c}\( \frac{r-r_c}{r_c-r_h}\)\d x^i \d\tau + \(\delta_{ij}- \frac{v_iv_j}{r_c^2 (1-v^2/r_c)} \(\frac{r-r_c}{1-r_h/r_c}\)\)\d x^i \d x^j\, .\label{eq:metric}
\end{align}
While this metric appears complicated, it is in reality simply flat space written in a complicated coordinate system.
Nevertheless, the Brown-York stress tensor on $\Sigma_c$ has the desired form \eqref{stresstensor}, where the energy density, pressure and relativistic fluid velocity are given by
\[
\rho = 0,\qquad p = \frac{1}{\sqrt{r_c-r_h}} ,\qquad u^a = \frac{1}{\sqrt{r_c-v^2}}(1,v_i) \label{pressequil}\, .
\]
In particular, the energy density $\rho$ vanishes.
We stress that there is no flat solution continuously connected to Rindler spacetime which generates a nonzero equilibrium energy density while at the same time preserving the form of the induced metric on $\Sigma_c$, as shown in appendix \ref{app:no_rho}.
In connection with this, note that the inherent ambiguity in the definition of the Brown-York stress tensor \cite{Brown:1992br},
which gives the freedom to send $T_{ab}\rightarrow T_{ab}+C\g_{ab}$ for some constant $C$, produces only a trivial constant shift in the background energy density and pressure given by $\rho \rightarrow \rho-C$ and $p\rightarrow p+C$.  This ambiguity has no dynamical consequences, however, and so we will suppress it in the remainder of this paper by setting $C=0$. Furthermore, since the induced metric $\g_{ab}$ is flat, there are no additional ambiguities involving the intrinsic curvature of $\Sigma_c$.

Finally, let us consider the thermodynamic properties of the metric \eqref{eq:metric}.
First, the Rindler horizon is a Killing horizon and has a constant Unruh temperature given by
\[
T = \frac{\kappa}{2\pi} = \frac{1}{4\pi \sqrt{r_c-r_h}},
\]
where the surface gravity $\kappa$ is defined as $\xi^\nu \nabla_\nu \xi^\mu = \kappa \xi^\mu$, where $\xi = \frac{1}{\sqrt{r_c -v^2}} (\partial_\tau + v_i \p_i)$ is normalised so that $\xi_a \xi^a = -1$ on $\Sigma_c$. If we then introduce an entropy density associated with the Rindler horizon identical to the black hole entropy density $s=1/4G$, we note the thermodynamic relation\footnote{The relation \eqref{eq:stp} may also be derived by evaluating Komar integrals both at the horizon $\Sigma_h$ and at $\Sigma_c$. Since the solution is stationary, homogeneous in $x^i$ and Ricci flat, the Komar integrals do not depend on the radius of evaluation and one can relate $sT$ defined at the horizon to $p$ defined on $\Sigma_c$ using standard manipulations, see e.g., \cite{Bardeen:1973gs,Townsend:1997ku}.}
\[
sT = p\, .\label{eq:stp}
\]
This further implies that two neighbouring equilibrium configurations are related by the Gibbs-Duhem relation
\[
s \delta T = \delta p \,.
\]
As a final point of interest, we note that the region $r_h \leq r \leq r_h + (1-r_h/r_c)v^2$ is an ergoregion with respect to the $\tau$-translations, which may lead to effects analogous to superradiance.

\section{Seed metric for near-equilibrium configurations}
\label{Se:seed}

Let us now change gears and discuss how to construct near-equilibrium configurations. We would like to obtain a solution of the vacuum Einstein equations parameterised by a velocity field $v_i$ and a pressure field $p$ which are local slowly varying functions of the coordinates $x^a$,
while preserving the form of the induced metric on $\Sigma_c$.
Working perturbatively in $\ep$, in the present section we show how to construct this solution up to order $\ep^2$.
In the following section, we then show how to extend this `seed' solution through to all orders in $\ep$.

Clearly, if we promote $v_i$ and $p$ to depend arbitrarily on the coordinates $x^a$, the metric \eqref{eq:metric} is no longer an exact solution of the vacuum Einstein equations. However, treating $v_i = v_i^{(\ep)}(\t,\vec{x}) $ and $p =r_c^{-1/2} + r_c^{-3/2}P^{(\ep)}(\t,\vec{x})$ as small fluctuations around the background $v_i = 0$, $p=r_c^{-1/2}$ in the hydrodynamic limit
\[
v_i^{(\ep)}(\t,\vec{x}) = \ep v_i(\ep^2\t,\ep\vec{x}), \qquad P^{(\ep)}(\t,\vec{x})=\ep^2 P(\ep^2\t,\ep \vec{x}),
\]
upon expanding \eqref{eq:metric} (noting that $r_h=2P^{(\ep)}+O(\ep^4)$) we obtain
\begin{align}
\label{seed}
\d s^2 &= -r \d\t^2+2\d\t\d r+\d x_i\d x^i \nn\\
&\quad -2\(1-\frac{r}{r_c}\)v_i\d x^i \d \t-\frac{2v_i}{r_c}\d x^i \d r \nn\\
&\quad +\(1-\frac{r}{r_c}\)\Big[(v^2+2P)\d\t^2+\frac{v_iv_j}{r_c}\d x^i\d x^j\Big]+\(\frac{v^2}{r_c}+\frac{2P}{r_c}\)\d\t\d r
+O(\ep^3),
\end{align}
where each successive line is of one order higher in the $\eps$ expansion.  (Here, and in the following, we drop the superscripts on $P^{(\ep)}$ and $v^{(\ep)}_i$ for simplicity.) This metric by construction
preserves the induced metric $\g_{ab}$ on $\Sigma_c$.  Moreover, it solves Einstein's equations to $O(\ep^3)$, provided one additionally imposes  incompressibility, i.e., $\D_iv^i = O(\ep^3)$. The metric \eqref{seed} reproduces the solution found in \cite{Bredberg} through to $\ep^2$ order.

As shown in \cite{Bredberg}, the Brown-York stress tensor on $\Sigma_c$ for the seed solution \eqref{seed} is
\begin{align}
\label{seed_stress}
T_{ab}\d x^a \d x^b &= \frac{\d \vec{x}^2}{\sqrt{r_c}}
- \frac{2 v_i}{\sqrt{r_c}}\,\d x^i\d \t
+\frac{v^2}{\sqrt{r_c}}\,\d\t^2 + r_c^{-3/2}\Big[P\delta_{ij}+v_iv_j-2r_c \D_i v_j\Big]\d x^i\d x^j +O(\eps^3)\, .
\end{align}
It is interesting to compare this result with the hydrodynamic stress tensor for a relativistic fluid with vanishing equilibrium energy density.
Up to first order in fluid gradients, this takes the form\footnote{As noted previously, and as we will discuss in section \ref{Section:Hydro}, for a fluid with vanishing equilibrium energy density the relativistic divergence $\K=\D_au^a$ vanishes at first order in gradients and so the usual bulk viscosity term is absent.}
\[
\label{std_hydroT}
T_{ab} = \rho u_a u_b + p h_{ab} -2 \eta  \K_{ab} + O(\D^2) ,
\]
where $\K_{ab} = h_a^c h_b^d\D_{(c}u_{d)}$ is the extrinsic curvature of surfaces orthogonal to the fluid velocity\footnote{We reserve the symbol $K_{ab}$ for the extrinsic curvature of $\Sigma_c$ itself.} and plays the role of the fluid shear.
Expanding this hydrodynamic stress tensor in $\ep$ to order $\ep^2$, one finds that one can indeed reproduce \eqref{seed_stress} upon setting
\[
\rho = 0 + O(\eps^3),\qquad p = \frac{1}{\sqrt{r_c}} + \frac{P}{r_c^{3/2}}+ O(\eps^3)\, ,\qquad \eta = 1\, .
\]

In the following section, we will now describe the algorithm to extend our seed solution \eqref{seed} to arbitrarily high order in $\eps$,
with the pressure and velocity perturbations obeying the incompressible Navier-Stokes equations with higher order corrections. Note that, if desired,
spacetimes describing fluctuations around a background with nonzero velocity and with a Rindler horizon located at
arbitrary $r=r_h$ may then be found by applying the finite diffeomorphisms \eqref{t1} and \eqref{t2} to the resulting solution.

\section{Constructing the solution to all orders}
\label{Section_algorithm}

Let us begin by assuming we know the bulk metric through order $\ep^{n-1}$.
The first nonvanishing components of the Ricci tensor  appear therefore at order $\ep^n$.
We now wish to add a new term  $g^\n_{\mu\nu}$ to the metric at order $\ep^n$ so as to extend our solution to order $O(\ep^{n+1})$.  In addition to adding a new term to the metric, we must also consider the effect of gauge transformations $\xi^{\n\mu}$ and field redefinitions $\delta v_i^\n$ and $\delta P^\n$ also at order $\ep^n$.

Noting that
\[
\D_r \sim \ep^0,  \qquad \D_i \sim \ep^1, \qquad \D_\t \sim \ep^2,
\]
adding a new term $g^\n_{\mu\nu}$ at order $\ep^n$ to the bulk metric produces a change in bulk Ricci tensor 

\noindent at the same order given by
\begin{align}
\label{linearRicci}
\delta R_{rr}^\n &= -\frac{1}{2}\D_r^2 g^\n_{ii},  \nn\\
\delta R_{ij}^\n &= -\frac{1}{2}\D_r(r\D_r g^\n_{ij}),  \nn\\
\delta R_{\t i}^\n &= -r\delta R_{ri}^\n = -\frac{r}{2}\D_r^2 g_{\t i}^\n, \nn\\
\delta R_{\t \t}^\n &=-r \delta R_{r \t}^\n = -\frac{r}{4} \(\D_r(r g^\n_{rr})+2\D_r g_{r\t}^\n-\D_r g_{ii}^\n+2\D_r^2 g_{\t\t}^\n\),
\end{align}
where we write $g_{ii}^\n \equiv \delta^{ij} g_{ij}^\n$ and $\delta R_{ii}^\n \equiv \delta^{ij}\delta R_{ij}^\n$.

Our goal then is to find a $g_{\mu\nu}^\n$ which cancels out the part of the Ricci tensor at order $\ep^n$ arising from the pre-existing metric up to order $\ep^{n-1}$.
(In fact, as we will see shortly, the boundary conditions are such that this requirement leads, after gauge-fixing, to a unique solution for
$g_{\mu\nu}^\n$.)
The full Ricci tensor at order $\ep^n$ must therefore satisfy
\[
\label{EinsteinEq}
R^{(n)}_{\mu\nu} = \delta R_{\mu\nu}^\n + \hat{R}_{\mu\nu}^\n = 0,
\]
where $\hat{R}_{\mu\nu}^\n$ denotes the part of the Ricci tensor at order $\ep^n$ arising from the metric up to order $\ep^{n-1}$.
Inspecting the equations for $\delta R_{\mu\nu}^\n$ above, we see that such a cancellation will only be possible provided the following integrability conditions are satisfied:
\[
\label{int_condns}
0 = \D_r(\hat{R}_{ii}^\n-r\hat{R}_{rr}^\n) -\hat{R}_{rr}^\n, \qquad
0 = \hat{R}_{\t a}^\n+r\hat{R}_{r a}^\n.
\]

To verify that these conditions are indeed satisfied, we
evaluate the Bianchi identity at order $\ep^n$ yielding\footnote{Note that the Ricci tensor $R_{\mu\nu}$ is itself of order $\ep^n$ (as we have already solved the vacuum Einstein equations at order $\ep^{n-1}$), so all the remaining covariant derivatives, metrics and inverse metrics appearing in the Bianchi identity are those of the background Rindler metric.}
\begin{align}
\label{Bianchi_id}
0 &= \D_r(\hat{R}_{ii}^\n-r\hat{R}_{rr}^\n) -\hat{R}_{rr}^\n, \nn\\
0 &= \D_r(\hat{R}_{\t a}^\n+r\hat{R}_{r a}^\n) \quad \Rightarrow \quad  \hat{R}_{\t a}^\n+r\hat{R}_{r a}^\n = f_a^\n(\t,\vec{x}).
\end{align}
The integrability conditions are therefore satisfied provided the arbitrary function $f_a^\n(\t,\vec{x})$ vanishes.

Evaluating the Gauss-Codazzi identity on $\Sigma_c$ at order $\ep^n$, we find
\[
\label{Gauss_Codacci}
\nabla^bT_{ab}\big|^\n_{\Sigma_c} = [2\nabla^b(K\g_{ab}-K_{ab})]^\n = [-2R_{a\mu}N^\mu]^\n =  -\frac{2}{\sqrt{r_c}}(\hat{R}_{\t a}^\n+r_c\hat{R}_{r a}^\n)=-\frac{2}{\sqrt{r_c}}f_a^\n(\t,\vec{x}).
\]
Thus, imposing the conservation of the Brown-York stress tensor on $\Sigma_c$ at order $\ep^n$ enforces the vanishing of $f_a^\n(\t,\vec{x})$, ensuring the validity of the integrability conditions \eqref{int_condns}.
From the perspective of the dual fluid, at order $\ep^3$ (i.e., the first order beyond the seed solution), this constraint reduces to the exact Navier-Stokes equation.  At subsequent even orders in $\ep$, this step amounts to adding corrections to the incompressibility condition, while at higher odd orders it amounts to adding corrections to the Navier-Stokes equation.

It is interesting to note
that evaluating the Hamiltonian constraint on $\Sigma_c$ yields an exact and nontrivial constraint on the Brown-York stress tensor of the dual fluid, namely
\[
\label{Ham_constraint}
d T_{ab}T^{ab} = T^2.
\]
Applying this constraint alone to equilibrium fluid configurations, one finds that either the equilibrium energy density vanishes $\rho_{\mathrm{eqm}}=0$ (as is the case here) or else is given by $\rho_{\mathrm{eqm}}=\left(-2d/(d-1)\right) p$, where $p$ is the equilibrium pressure. It would be interesting to see if this latter solution branch could be obtained through a modification of the present scenario. We will discuss further the implications of the constraint \eqref{Ham_constraint} for the dual hydrodynamics in section \ref{Section:Hydro}.

We may now proceed to use \eqref{linearRicci} to identify a particular solution $\tilde{g}^\n_{\mu\nu}$ that satisfies the Einstein equation \eqref{EinsteinEq} at order $\ep^n$.  One such particular solution is
\begin{align}
\tilde{g}_{r\mu}^\n &=  0,\nn\\
\tilde{g}_{\t\t}^\n &= \beta_1^\n(\t,\vec{x})+(1-r/r_c)\beta_2^\n(\t,\vec{x})+\int_r^{r_c}\d r'\int_{r'}^{r_c}\d r'' (\hat{R}_{ii}^\n-r\hat{R}_{rr}^\n-2\hat{R}_{r\t}^\n),\nn\\
\tilde{g}_{\t i}^\n &= \beta_{3i}^\n(\t,\vec{x})+(1-r/r_c)\beta_{4i}^\n(\t,\vec{x})-2\int_r^{r_c}\d r'\int_{r'}^{r_c}\d r'' \hat{R}_{ri}^\n,\nn\\
\tilde{g}_{ij}^\n&=\beta_{5ij}^\n(\t,\vec{x})+\ln(r/r_c)\beta_{6ij}^\n(\t,\vec{x})-2\int^{r_c}_r\d r' \frac{1}{r'}\int^{r'}_{r_*}\d r'' \hat{R}_{ij}^\n.
\end{align}
Here, we have chosen limits so that the integrals vanish for $r=r_c$, facilitating the evaluation of boundary conditions to follow shortly.
We will leave the lower limit $r_*$ appearing in the inner integral for $\tilde{g}_{ij}^\n$ arbitrary for now.
Note however that the $rr$ Einstein equation, in combination with the Bianchi identity \eqref{Bianchi_id}, fixes the trace $\beta_{6ii}^\n=\delta^{ij}\beta_{6ij}^\n$ to be
\[
\label{beta_constraint}
\beta_{6ii}^\n(\t,\vec{x}) = 2r_*(\hat{R}_{ii}^\n-r_*\hat{R}_{rr}^\n)\big|_{r=r_*}.
\]

Allowing now for gauge transformations $\xi^{\n\mu}(r,\t,\vec{x})$ at order $\ep^n$, as well as field redefinitions $\delta v_i^\n(\t,\vec{x})$ and $\delta P^\n(\t,\vec{x})$ also at this order, the above solution generalises to
\begin{align}
\label{gsoln}
g_{rr}^\n &= 2\D_r\xi^{\n\t}, \nn\\
g_{r\t}^\n &= -r\D_r\xi^{\n\t}+\D_r\xi^{\n r}+\frac{1}{r_c}\delta P^\n,\nn\\
g_{ri}^\n &= \D_r\xi_i^\n-\frac{1}{r_c}\delta v_i^\n,\nn\\
g_{\t\t}^\n &= \tilde{g}_{\t\t}^\n-\xi^{\n r}+(1-r/r_c)2\delta P^\n,\nn\\
g_{\t i}^\n &= \tilde{g}_{\t i}^\n-2(1-r/r_c)\delta v_i^\n, \nn\\
g_{ij}^\n &= \tilde{g}_{ij}^\n.
\end{align}
Here, to obtain the terms involving $\delta v_i^\n$ and $\delta P^\n$ it is sufficient to look at
how $v_i$ and $P$ appear linearly in the seed solution \eqref{seed}, since $\delta v_i^\n \sim \delta P^\n \sim \ep^n$.
Note that the $g_{\mu\nu}^\n$ above is still a solution of Einstein's equations, since the additional terms 
coming from field redefinitions and gauge transformations  do not contribute to $\delta R_{\mu\nu}^\n$, as we see from \eqref{linearRicci}.

To fix the gauge, we will impose $g^\n_{r\mu}=0$, for all $n>2$.  This choice is convenient since we do not have boundary conditions at $\Sigma_c$ for these metric components, unlike for the $g^\n_{ab}$.
The $g_{r\mu}$ components of the full metric are then simply those of the seed solution, i.e., to all orders in $\ep$,
\[
\label{gr_components}
 g_{rr} = 0, \qquad g_{r\t} =1+\frac{v^2}{2r_c}+\frac{P}{r_c} ,\qquad g_{ri} = -\frac{v_i}{r_c}.
\]
Note this gauge choice is fully consistent with our considerations in appendix \ref{app:radialgauge} (cf.~equation \eqref{gauge1}).
To impose this gauge choice then, we must fix
\begin{align}
& \xi^{\n r} = (1-r/r_c)\delta P^\n +\tilde{\xi}^{\n r}(\t,\vec{x}), \qquad \xi^{\n\t} = \tilde{\xi}^{\n\t}(\t,\vec{x}),\nn\\[1ex]
&\qquad\qquad \xi^\n_i = -(1-r/r_c)\delta v_i^\n+\tilde{\xi}^\n_i(\t,\vec{x}),
\end{align}
where the $\tilde{\xi}^{\n\mu}(\t,\vec{x})$ parameterise the residual gauge freedom.
Substituting this back into \eqref{gsoln}, we find
\begin{align}
\label{newgsoln}
g_{r\mu}^\n &= 0, \nn\\
g_{\t\t}^\n &=  \beta_1^\n-\tilde{\xi}^{\n r}+(1-r/r_c)(\beta_2^\n+\delta P^\n)+\int_r^{r_c}\d r'\int_{r'}^{r_c}\d r'' (\hat{R}_{ii}^\n-r\hat{R}_{rr}^\n-2\hat{R}_{r\t}^\n),\nn\\
g_{\t i}^\n &= \beta_{3i}^\n+(1-r/r_c)(\beta_{4i}^\n-2\delta v_i^\n)-2\int_r^{r_c}\d r'\int_{r'}^{r_c}\d r'' \hat{R}_{ri}^\n, \nn\\
g_{ij}^\n &= \beta_{5ij}^\n+\ln(r/r_c)\beta_{6ij}^\n-2\int^{r_c}_r\d r' \frac{1}{r'}\int^{r'}_{r_*}\d r'' \hat{R}_{ij}^\n.
\end{align}

Imposing the boundary condition $g^\n_{ab}=0$ on $\Sigma_c$ (so that the induced metric $\g_{ab}$ remains fixed) then fixes
\[
\label{beta_fix}
 \beta_1^\n(\t,\vec{x}) = \tilde{\xi}^{\n r}(\t,\vec{x}), \qquad \beta_{3i}^\n(\t,\vec{x})=0, \qquad  \beta_{5ij}^\n(\t,\vec{x})=0.
\]
To fix the trace-free part of $\beta_{6ij}^\n$, we require that the metric must be regular at the future horizon $r=0$, or equivalently analytic, at each order $\ep^n$.
To achieve this, it is useful to choose the lower limit $r_*$ in the inner integral for $g_{ij}^\n$ in \eqref{newgsoln} to be zero.
Since $\hat{R}_{ij}^\n$ will turn out to be a polynomial in $r$, by setting $r_*=0$ we ensure that the outer integral does not generate any additional logarithmic terms beyond those associated with the $\beta_{6ij}^\n$.  (Moreover, the trace part of $\beta_{6ij}^\n$ is set to zero through \eqref{beta_constraint}.)
Regularity on the future horizon at order $\ep^n$ then requires\footnote{A possible caveat here is that we impose
order by order regularity but the logarithmic terms could yield a finite value on the future horizon when resummed to all orders in $\ep$. We thank Bob Wald for discussion on this point.}
that these remaining logarithmic terms vanish:
\[
 \beta_{6ij}^\n(\t,\vec{x})=0.
\]

In conclusion then, with these boundary conditions, the new part of the bulk metric at order $\ep^n$ is therefore
\begin{align}
\label{finalsoln}
g_{r\mu}^\n &= 0, \nn\\
g_{\t\t}^\n &=  (1-r/r_c)F_\t^\n(\t,\vec{x})+\int_r^{r_c}\d r'\int_{r'}^{r_c}\d r'' (\hat{R}_{ii}^\n-r\hat{R}_{rr}^\n-2\hat{R}_{r\t}^\n),\nn\\
g_{\t i}^\n &= (1-r/r_c)F_i^\n(\t,\vec{x})-2\int_r^{r_c}\d r'\int_{r'}^{r_c}\d r'' \hat{R}_{ri}^\n, \nn\\
g_{ij}^\n &= -2\int^{r_c}_r\d r' \frac{1}{r'}\int^{r'}_{0}\d r'' \hat{R}_{ij}^\n,
\end{align}
where the arbitrary functions are
\[
\label{F_relations}
 F_\t^\n(\t,\vec{x}) = \beta_2^\n(\t,\vec{x})+\delta P^\n(\t,\vec{x}), \qquad F_i^\n(\t,\vec{x}) = \beta_{4i}^\n(\t,\vec{x})-2\delta v_i^\n(\t,\vec{x}).
\]
We will discuss gauge choices to fix these arbitrary functions in the following section.

As emphasized above,
at each order $\ep^n$ this integration scheme generates a solution that is regular on the future horizon $r=0$.
If $\hat{R}_{\mu\nu}^\n$ is a polynomial in $r$ that is regular at $r=0$, then $g_{\mu\nu}^\n$ will itself be regular at $r=0$. This in turn implies that at the next order $\hat{R}_{\mu\nu}^{(n+1)}$ will be regular as well. Since for the seed solution $\hat{R}_{\mu\nu}^{(3)}$ is regular, this argument ensures regularity of $g_{\mu\nu}^\n$ at $r=0$ for each $n$.

An additional
interesting feature of the hydrodynamic expansion is fact that any vector constructed from $v_i$, $P$ and their derivatives is necessarily of odd order in $\ep$, while any scalar or rank-two tensor constructed from these variables must necessarily be of even order in $\ep$.
This means that at orders $\ep^n$, where $n$ is odd, the Ricci tensor components $\hat{R}_{rr}^\n$, $\hat{R}_{\t\t}^\n$ and $\hat{R}_{ij}^\n$ and the arbitrary function $F_\t^\n(\t,\vec{x})$ all vanish identically, and so the only nonzero metric component generated by the integration scheme \eqref{finalsoln} is $g_{\t i}^\n$.  Correspondingly, at orders $\ep^n$ where $n$ is now even, $\hat{R}_{\t i}^\n$ and $F_i^\n(\t,\vec{x})$ are zero, and so and so the only nonzero metric components generated by the integration scheme are $g_{\t\t}^\n$ and $g_{ij}^\n$.

\section{Gauge choices for the fluid}
\label{gauge-choice}

In this section we discuss the gauge choice for the fluid so as fix the arbitrary functions $F_a^\n(\t,\vec{x})$ appearing in the integration scheme \eqref{finalsoln}.  In preparation, let us begin by evaluating the contribution $\delta T_{ab}^\n$ to the Brown-York stress tensor arising from the new term $g_{\mu\nu}^\n$ added to the metric at order $\ep^n$.

The variation in the extrinsic curvature of $\Sigma_c$ at order $\ep^n$ due to $g_{\mu\nu}^\n$ is
\[
\delta K^{(n)}_{ab} = \frac{1}{2}\pounds_{N} g_{ab}^\n = \frac{1}{2}N^r\D_r g_{ab}^\n = \frac{1}{2}\sqrt{r_c}\D_r g_{ab}^\n\big|_{\Sigma_c},
\]
where, since $g_{\mu\nu}^\n$ is already of order $\ep^n$, only derivatives with respect to $r$ survive, and the normal is that associated with the background Rindler spacetime.
Thus, we have
\[
\delta K_{\t\t}^\n = -\frac{F_\t^\n(\t,\vec{x})}{2\sqrt{r_c}}, \qquad \delta K_{\t i}^\n = -\frac{F_i^\n(\t,\vec{x})}{2\sqrt{r_c}},
\qquad \delta K^\n_{ij} =  +\frac{1}{\sqrt{r_c}}\int_0^{r_c}\d r' \hat{R}_{ij}^\n.
\]

The new term due to $g_{\mu\nu}^\n$ in the Brown-York stress tensor on $\Sigma_c$ at order $\ep^n$ is then
\[
 \delta T_{ab}^\n = 2(\g_{ab}\delta K^\n-\delta K_{ab}^\n),
\]
which evaluates to
\begin{align}
\label{deltaT}
& \delta T_{\t\t}^\n = -2\sqrt{r_c}\int_0^{r_c}\d r' \hat{R}_{ij}^\n, \qquad \delta T_{\t i}^\n = \frac{F_i^\n(\t,\vec{x})}{\sqrt{r_c}}, \nn\\[1ex]
&\delta T_{ij}^\n = \frac{F_\t^\n(\t,\vec{x})}{r_c^{3/2}}\,\delta_{ij}+\frac{2}{\sqrt{r_c}}\int_0^{r_c}\d r' (\delta_{ij}\hat{R}_{kk}^\n-\hat{R}_{ij}^\n).
\end{align}

The complete Brown-York stress tensor on $\Sigma_c$ at order $\ep^n$ is then
\[
 T^{(n)}_{ab} = \delta T_{ab}^\n + \hat{T}_{ab}^\n,
\]
where $\hat{T}_{ab}^\n$ represents the contribution at order $\ep^n$ due to the metric up to order $\ep^{n-1}$.

\subsection{Fixing \texorpdfstring{$F_i^\n(\t,\vec{x})$}{Fi(t,x)}}

From \eqref{F_relations}, we see that the arbitrary vector function $F_i^\n(\t,\vec{x})$ appearing at odd orders in $\ep$ is related to redefinitions of the fluid velocity.  To fix this ambiguity, we define the fluid velocity as the boost from the lab frame to the local rest frame of the fluid, in which the momentum density vanishes
(i.e., $T_{\t i}=0$ in the local rest frame where the fluid velocity is purely timelike).
In an arbitrary frame where the fluid velocity $u^a=\gamma (1, v_i)$, this Landau gauge condition \cite{Landau} then reads
\[
\label{rel_no_mom}
0 = h_a^b T_{bc} u^c, \qquad h^a_b=\delta^a_b + u^a u_b.
\]
At odd orders in $\ep$, the $\t$ component of this equation vanishes identically, since there are no scalars at odd order. The remaining vector component reads
\[
\label{no_momentum}
0 = T_{i\t}^\n+T_{ij}^{(n-1)} v_j + \rho^{(n-1)} v_i,
\]
where $\rho=T_{ab}u^a u^b$ is the energy density in the local rest frame\footnote{While this quantity vanishes for the background solution, it is nonzero at higher order in fluid gradients.}.
Hence at order $\ep^n$ (for odd $n$), we find $F_i^\n(\t,\vec{x})$ is fixed in terms of known quantities according to
\[
\label{Fi_gauge}
 0 = \frac{F_i^\n(\t,\vec{x})}{\sqrt{r_c}}+\hat{T}^{(n)}_{i\t}+T^{(n-1)}_{ij}v_j + \rho^{(n-1)}v_i,
\]
where $T^{(n-1)}_{ij}$ and $\rho^{(n-1)}$ denote respectively the full stress tensor and the energy density evaluated at order $\ep^{n-1}$, both of which are known quantities.

If we now move to order $\ep^{n+1}$ (where $n{+}1$ is even), we find the vector component of \eqref{rel_no_mom} vanishes identically (as there are no vectors at even orders), but the $\t$ component is nontrivial and reads
\[
\label{consistency_check}
 0 = T^{(n+1)}_{\t\t}+T_{\t i}^{(n)}v_i-r_c\rho^{(n+1)}.
\]
Noting that $\rho = \gamma^2(T_{\t\t}+2T_{\t i}v_i+T_{ij}v_iv_j)$ and that $1-r_c\gamma^2 = -v^2\gamma^2$, one can show that \eqref{consistency_check} is equivalent to
\[
 0 = v^2 (T_{\t\t}^{(n-1)}+T_{\t i}^{(n-2)}v_i) + r_c(T_{\t i}^{(n)}+T_{ij}^{(n-1)}v_j)v_i .
\]
Then, assuming \eqref{no_momentum} is satisfied at order $\ep^{n}$, \eqref{consistency_check} reduces to
\[
 0 = v^2 (T_{\t\t}^{(n-1)}+T_{\t i}^{(n-2)}v_i - r_c \rho^{(n-1)}).
\]
Thus, \eqref{consistency_check} is satisfied at order $\ep^{n+1}$ provided that it is satisfied at the preceding even order $\ep^{n-1}$.  Since the seed solution \eqref{seed} explicitly satisfies \eqref{consistency_check} at order $\ep^2$, we see that the gauge condition \eqref{rel_no_mom} can indeed be consistently imposed at all orders by choosing $F_i^\n(\t,\vec{x})$ according to \eqref{Fi_gauge}.

It is quite nontrivial that we can consistently impose the {\it relativistic} gauge choice \eqref{rel_no_mom} on the dual fluid, since the latter is constructed according to a non-relativistic expansion scheme in which time and spatial derivatives have different order in $\ep$.  We will return to this topic in section \ref{Section:Hydro}.

\subsection{Fixing \texorpdfstring{$F_\t^\n(\t,\vec{x})$}{Ft(t,x)}}
\label{Section:Ft}

From \eqref{F_relations}, we see that $F_\t^\n(\t,\vec{x})$, the arbitrary scalar function appearing at all even orders in $\ep$ greater than two, is related to redefinitions of the pressure fluctuation $P$.
To remove this ambiguity, we propose defining the pressure fluctuation $P$ so that the isotropic part of $T_{ij}$ is fixed to be
\[
\label{Ft_gauge}
 T^{\mathrm{isotropic}}_{ij} = \(\frac{1}{\sqrt{r_c}}+\frac{P}{r_c^{3/2}}\)\delta_{ij}
\]
to all orders.
From \eqref{deltaT}, it is clear that we can always choose $F_\t^\n(\t,\vec{x})$ at every even order greater than two so as to ensure that $T_{ij}$ receives no higher order corrections that are proportional to $\delta_{ij}$.

While other gauge choices to define the pressure fluctuation $P$ are possible (see footnote \ref{gaug_ch}), the present choice is especially simple from an operational point of view.

\section{Results to \texorpdfstring{$O(\eps^6)$}{fifth order}}
\label{explicit}

In this section we now apply the integration scheme developed in the preceding sections to compute the bulk metric and the ensuing fluid stress tensor at orders up to and including $\ep^5$ in four and five bulk spacetime dimensions. We have checked explicitly at the end of the procedure that our solution satisfies the bulk Einstein equations to $O(\ep^6)$ for four and five dimensions.

Since neither the seed metric, nor the algorithm to construct the solution at each step, depends on the dimension; and moreover
scalars, vectors and rank-two tensors formed from $v_i$, $P$ have no special properties in three spatial dimensions
(unlike in two spatial dimensions\footnote{In two spatial dimensions (i.e., $d=2$), $v_i$ has two components and dimension dependent identities for rank 2 tensors exist at order $\eps^4$. Notably, $\s_{ik}\s_{kj}=1/2 \delta_{ij}\s_{kl}\s_{kl}$ and $\omega_{ik}\omega_{kj}=1/2 \delta_{ij}\omega_{kl}\omega_{lk}$, where $\s_{ij}$ and $\omega_{ij}$ are as defined in \eqref{sigmadef}.}),
we expect that the following results are valid in \emph{any} bulk dimension greater than five as well. We have not, however, explicitly checked that the equations of motion hold to $O(\ep^6)$ for bulk dimensions greater than five.

\subsection{Corrections to Navier-Stokes and incompressibility}

As noted in \cite{Bredberg}, imposing conservation of the Brown-York stress tensor at order $\ep^2$ yields the incompressibility condition, while at order $\ep^3$ we recover the Navier-Stokes equation.  In section \ref{Section_algorithm}, we saw explicitly how conservation of the Brown-York stress tensor at order $\ep^n$ is required in order to construct the bulk metric at that same order.

Applying the integration scheme described in section \ref{Section_algorithm}, at order $\ep^4$ we obtain the following corrections to the incompressibility equation:
\[
\label{final_I}
\p_i v_i =  \frac{1}{r_c} v_i \p_i P -v_i \p^2 v_i +\frac{1}{2}\s_{ij}\s_{ij} + O(\ep^6),
\]
where the fluid shear $\s_{ij}$ and vorticity $\omega_{ij}$ are given by
\[
\label{sigmadef}
 \sigma_{ij} = 2\D_{(i}v_{j)} \equiv \D_iv_j+\D_jv_i, \qquad \omega_{ij} = 2\D_{[i}v_{j]} \equiv \D_iv_j-\D_jv_i.
\]
As \eqref{final_I} is a scalar equation, there are no corrections at odd orders, and so the next corrections appear at order $\ep^6$.

As expected, the Navier-Stokes equations are modified at order $\ep^5$:
\begin{align}
\p_\tau v_i +v_j \p_j v_i -r_c \p^2 v_i +\p_i P &= -\frac{3r_c^2}{2}\p^4 v_i+2r_cv_k \p^2 \p_k v_i+r_c \s_{ik}\p_l \s_{kl}-\frac{5r_c}{2}\omega_{ik}\p_l \s_{kl} \nn \\
&\quad -\frac{3r_c}{4}\p_i (\s_{kl}\s_{kl}) - \frac{5r_c}{8}\p_i (\omega_{kl}\omega_{lk})+r_c \s_{kl} \p_k \s_{li}-2v_k \p_k \p_i P \nn \\
&\quad -2 (\p_k v_i)\p_k P-P\p^2 v_i  - \frac{1}{2}v^2 \p^2 v_i - \frac{1}{2}(\p_k \s_{il})v_k v_l+\frac{1}{2}(\p_k \omega_{il}) v_k v_l \nn \\
&\quad +2(\p_k v_i )\omega_{kl}v_l + \frac{1}{r_c}(P+v^2) \p_i P - \frac{1}{r_c}v_i\p_\tau P   + O(\eps^7)\, .\label{final_NS}
\end{align}
Since this is a vector equation, there are no corrections at even orders, hence the next corrections will be at order $\ep^7$.

In writing the form of these correction terms, we eliminated $\D^2 P$ terms using the relationship
\[
\label{derived_eqn}
 \D^2 P + (\D_iv_j)(\D_jv_i)=O(\ep^6),
\]
which follows from taking the divergence of Navier-Stokes equation and using incompressibility.

\subsection{Metric at higher orders}

The initial seed metric up to $\ep^2$ is given in \eqref{seed}. Applying our solution algorithm for the bulk geometry, at order $\ep^3$ the only nonvanishing components of the metric are
\[ \label{g3}
 g^{(3)}_{\t i} = \frac{(r-r_c)}{2r_c}\Big[(v^2+2P)\frac{2v_i}{r_c}+4\D_iP -(r+r_c)\D^2 v_i\Big].
\]
Note that only the last of these terms was found in \cite{Bredberg}; the first two terms are required in order to impose the fluid gauge condition \eqref{no_momentum} at order $\ep^3$.

At $\ep^4$, the only nonvanishing components of the metric are
\[ \label{g41}
g^{(4)}_{\tau\tau} = -\frac{(r-r_c)^3}{8 r_c^2}(\omega_{kl}\omega_{lk}) +\frac{(r-r_c)^2}{8 r_c}( 8 v_k\p^2 v_k + \s_{kl}\s_{kl})-\frac{(r-r_c)}{r_c} F_{\tau}^{(4)},
\]
where
\[
F^{(4)}_{\tau} = \frac{9}{8r_c}v^4+\frac{5}{2r_c}Pv^2+\frac{P^2}{r_c}-2r_c v_i \p^2 v_i - \frac{r_c}{2}\s_{kl}\s_{kl}-2\partial_\tau P + 2v_k \partial_k P,
\]
is fixed by the gauge condition. Also,
\begin{align}
g^{(4)}_{ij} = &(1-\frac{r}{r_c}) \Big[ \frac{1}{r_c^2}v_i v_j (v^2+2P)+\frac{2}{r_c}v_{(i}\p_{j)}P-4\p_i\p_j P -\frac{1}{2}\s_{ik}\s_{kj}+\frac{r-5r_c}{4r_c}\omega_{ik}\omega_{kj} \nonumber\\
&\quad + \s_{k(i}\omega_{j)k}-\frac{r+r_c}{r_c}v_{(i}\p^2 v_{j)}+\frac{r+5r_c}{4}\p^2\s_{ij}-\frac{1}{r_c}v_{(i}\p_{j)}v^2-\frac{1}{2r_c}\s_{ij}(v^2+2P)\Big]. \label{g42}
\end{align}

At next order we find that the only nonvanishing components of the metric are
\begin{align} \label{g5}
g^{(5)}_{\tau i} =(1-\frac{r}{r_c}&)  \Big[ -\frac{(r+2r_c)(r+5r_c)}{12}\p^4 v_i + \frac{r-3r_c}{4}v_k \p^2 \s_{ik}-\frac{1}{2}(r+r_c)v_k \p^2 \omega_{ik}+r_c \s_{ik}\p_l \s_{kl} \nn \\
&\quad - \frac{(r+r_c)(r+4r_c)}{4r_c}\omega_{ik}\p_l \s_{kl}-\frac{5r+r_c}{8}\p_i (\s_{kl} \s_{kl})+\frac{r^2-7r_c r-4 r_c^2}{16 r_c}\p_i (\omega_{kl}\omega_{lk}) \nn \\
&\quad +\frac{r+r_c}{2}\s_{kl}\p_k \s_{li} - \frac{r-3r_c}{r_c}v_k \p_k \p_i P - \s_{ik}\p_k P + \frac{r}{r_c}\omega_{ik}\p_k P + \frac{r-2r_c}{r_c}\p^2 v_i P  \nn \\
 &\quad + \frac{r}{2r_c}v^2 \p^2 v_i + \frac{r}{r_c}v_i v_l \p_k \s_{kl} - \frac{r+r_c}{4r_c}\p_k \s_{il}v_k v_l + \frac{r+r_c}{4r_c}\p_k \omega_{ik}v_k v_l + \frac{1}{2}\s_{ik}\s_{kl}v_l \nn \\
 &\quad  + \frac{r+2r_c}{2r_c} \s_{ik} \omega_{kl}v_l - \frac{1}{2}\omega_{ik}\s_{kl}v_l - \frac{3r-3r_c}{4r_c}\omega_{ik}\omega_{kl}v_l + \frac{r+3r_c}{8r_c}\s_{kl}\s_{kl}v_i \nn \\
 &\quad - \frac{(r-r_c)^2}{8r_c^2}\omega_{kl}\omega_{lk}v_i - \frac{1}{r_c^2}P^2 v_i - \frac{5}{2r_c^2}P v_i v^2 - \frac{9}{8r_c^2}v_i (v^2)^2+\frac{1}{r_c}P \p_i P  \nn \\
 &\quad  + \frac{r-r_c}{r_c^2}\p_i v^2 P- \frac{1}{r_c}v_k \p_k P v_i - \frac{r-2r_c}{r_c^2}v_k \s_{ki}P - \frac{1}{r_c}\p_\tau P v_i + \frac{1}{r_c}\s_{ik}v_k v^2 \nn\\
 &\quad + \frac{r}{2r_c^2}\omega_{ik}v_k v^2 + \frac{1}{2r_c}v_l \s_{kl}v_l v_i
\Big]\, .
\end{align}

Inspecting these results, we see that the bulk metric takes the form of a polynomial in $r$, and as such we would expect it to have only a finite radius of convergence centered about $r=r_c$. 
Note however that this location is arbitrary: by applying the scaling $(r,\tau,x_i,v_i,P)\rightarrow (\lambda^2 r, \tau, \lambda x_i, \lambda v_i, \lambda^2 P)$ we may set $r_c$ to any chosen value.
Let us also mention that, while we verified that the $d=4$ solution is Petrov II up to order $\eps^{12}$, as noted in \cite{Bredberg}, the metric fails to be of Petrov II type at order $\eps^{14}$.

Finally, let us comment on the location of the Rindler horizon for the full solution. This can be 
worked out perturbatively in the non-relativistic expansion, as in the analogous computation in \cite{Bhattacharyya:2008xc}.
Let $r=R(\tau,x_i)$ be this position, where $R$ is constructed from $v_i$, $P$ and their derivatives.  
The terms without any derivatives follow from the equilibrium solution, which yields $R_{eqm} = r_c [1-(1+P/r_c)^{-2}]$.  The remaining gradient terms may then be determined by imposing at each order in $\epsilon$ that the hypersurface be null: $0=g^{\mu\nu}\partial_\mu (r-R)\partial_\nu (r-R)$.  To $O(\ep^4)$, there are no gradient terms that can be constructed, hence $R=2P+O(\ep^4)$.  To obtain the non-trivial gradient terms appearing at $\epsilon^4$ order, however, would require knowing $g^{rr}$ to $\epsilon^6$ order.

\subsection{Stress tensor at higher orders}

The stress tensor of the seed metric was given in \eqref{seed_stress}. At order $\eps^3$, this stress tensor gets corrected by the following expression
\begin{align}
T^{(3)}_{ab}\d x^a \d x^b =2 r_c^{-3/2}\Big[r_c\s_{ik}v_k-(v^2+P)v_i\Big]\d x^i\d\t \, .\label{T3}
\end{align}
At order $\eps^4$, the only nonvanishing components of the stress tensor that receive a correction are the scalar $T^{(4)}_{\tau \tau}$ and tensor parts $T^{(4)}_{ij}$ as follows,
\begin{align}
T^{(4)}_{ab}\d x^a \d x^b &= r_c^{-3/2}\Big[v^2(v^2+P)-\frac{r_c^2}{2}\s_{ij}\s_{ij}-r_c \s_{ij}v_iv_j\Big]\d\t^2\nn \\[1ex]
&\quad  + r_c^{-5/2}\Big[ v_iv_j (v^2+P) +2r_c v_{(i}\p_{j)}P -4 r_c^2 \p_i\p_j P-\frac{r_c^2}{2}\s_{ik}\s_{kj}-r_c^2 \omega_{ik}\omega_{kj} \nn\\[1ex]
&\quad +r_c^2\s_{k(i}\omega_{j)k} -2r_c^2 v_{(i}\p^2 v_{j)}+\frac{3r_c^3}{2}\p^2 \s_{ij}-r_c v_{(i}\p_{j)}v^2 -\frac{r_c}{2}\s_{ij}v^2  \Big] \d x^i\d x^j \, .\label{T4}
\end{align}
At order $\ep^5$, we obtain
\begin{align}
T^{(5)}_{ab}\d x^a \d x^b &= 2 r_c^{-5/2}\Big[ -\frac{3r_c^3}{2}v_k \p^2 \s_{ik} +4 r_c^2 v_k \p_k \p_i P + r_c^2 v^2 \p^2 v_i
+r_c^2 v_i v_l \p_k \s_{kl} + \frac{r_c^2}{2}\s_{ik}\p_k v^2 \nn \\
& \quad - \frac{r_c^2}{2}\omega_{ik} \s_{kl}v_l +r_c^2 \omega_{ik}\omega_{kl}v_l + \frac{r_c^2}{2}\s_{kl}\s_{kl}v_i -  v_i v^2 (P+v^2)-r_c v^2\p_i P \nn \\
&\quad - r_c v_i v_k \p_k P +r_c \s_{ik}v_k v^2+\frac{r_c}{2}\omega_{ik}v_k v^2 + \frac{r_c}{2}v_k \s_{kl}v_l v_i  \Big] d\tau dx^i \, .\label{T5}
\end{align}

To simplify the form of these expressions, we have made use of the constraint equations \eqref{final_I} and \eqref{final_NS} below in such a manner that all $\tau$ derivatives of $v_i$ do not appear in the final form of the expressions, as well as equations deriving from these such as \eqref{derived_eqn} in such a manner that $\D^2 P$ does not appear in the final form of the expressions.

\section{Characterising the dual fluid}
\label{Section:Hydro}

Earlier, we noted how the Brown-York stress tensor of the seed metric could be obtained from the $\ep$-expansion of the hydrodynamic stress tensor for a relativistic fluid.
Moreover, when constructing the bulk solution, we found one can consistently choose a gauge such that $u^aT_{ab}h^c_c=0$ to all orders in $\ep$.
Taken together, these observations suggest that it might in fact be possible to recover our full gravitational results for the Brown-York stress tensor from
the $\ep$-expansion of some appropriately chosen relativistic hydrodynamic stress tensor.

Examining the results of the previous section more closely, we find that the energy density in the local rest frame is given by
\[ \label{rho_lead}
\rho = T_{ab}u^au^b = -\frac{1}{2\sqrt{r_c}}\s_{ij}\s_{ij}+O(\eps^6).
\]
In particular, we see that the energy density vanishes for equilibrium configurations in which the fluid velocity is everywhere constant.
In the following section, we discuss the theory of relativistic hydrodynamics at second order in fluid gradients, paying special attention to the modifications necessitated by the vanishing equilibrium energy density.
We will then proceed to match our proposed relativistic hydrodynamic stress tensor with the Brown-York stress tensor derived from our gravitational calculations above, permitting the identification of second-order transport coefficients.

\subsection{Relativistic hydrodynamics for vanishing equilibrium energy density}

Defining the relativistic fluid velocity $u^a$ as in \eqref{rel_no_mom}, so that the momentum density in the local rest frame vanishes,
the fluid stress tensor takes the general form
\[
\label{hydroT}
 T_{ab} = \rho u_a u_b + p h_{ab} + \Pi^\perp_{ab}, \qquad u^a\Pi^\perp_{ab}=0.
\]
Here, $h_{ab} = \g_{ab}+u_au_b$ is the induced metric on surfaces orthogonal to the fluid velocity, and $p$ represents the pressure of the fluid in the local rest frame.  The term $\Pi^\perp_{ab}$ encodes dissipative corrections and may be expanded in gradients of the fluid velocity.
Since in the present case the equilibrium energy density vanishes, we must also expand the energy density $\rho$ in terms of gradients of the fluid velocity.
Inserting \eqref{hydroT} into the constraint \eqref{Ham_constraint} deriving from the bulk Hamiltonian constraint, we find
\[
\label{rho_result}
0 = \rho \big( (d-1)\rho+2d p+2\Pi^{\perp}\big) + d\Pi^\perp_{ab}\Pi^{\perp ab}-(\Pi^\perp)^2,
\]
where $\Pi^\perp = h^{ab}\Pi^\perp_{ab}$.
This relation then fully determines the energy density $\rho$ in terms of $p$ and $\Pi^\perp_{ab}$ (remembering that we must select the solution branch corresponding to a zero equilibrium energy density). Equation \eqref{rho_result}
therefore plays a role somewhat analogous to that of the equation of state for a conventional fluid.

To write down the gradient expansions for $\rho$ and $\Pi^\perp_{ab}$ precisely, it is useful to first consider the equations of motion at lowest order in fluid gradients: noting that $\rho$ has no component at zeroth order in gradients, these are
\[
\label{zerograd}
 0 = u^a\D^b T_{ab} = -p \D_a u^a + O(\D^2), \qquad 0=h_a^b\D^cT_{bc} \quad \Rightarrow\quad D u_a =- D^\perp_a \ln p +O(\D^2),
\]
where we have omitted terms of second and higher order in fluid gradients, and we have defined $D^\perp_a \equiv h_a^b\D_b$ and $D \equiv u^a\D_a$.  We may use these equations to simplify the form of the possible coefficients that appear at a given order in the gradient expansion: for example, we see that the only possible first order term in the gradient expansion for $\rho$, namely $\D_a u^a$, in fact vanishes at this order.  The expansion for $\rho$ therefore starts at second order.

As shown in appendix \ref{hydro_app}, making use of the relations \eqref{zerograd} we may write down the following complete basis for $\Pi^\perp_{ab}$ up to second order in gradients,
\begin{align}
\label{PiT}
\Pi^\perp_{ab} &= -2\eta \K_{ab} + c_1 \K_a^c\K_{cb} + c_2 \K_{(a}^c\Omega_{|c|b)} + c_3 \Omega_a^{\,\,\,c}\Omega_{cb} +
c_4 h_a^ch_b^d\D_c\D_d \ln p \nn\\
& \quad  + c_5 \K_{ab}\,D\ln p + c_6 D^\perp_a \ln p \,D^\perp_b\ln p + O(\D^3),
\end{align}
where $\eta$ is the relativistic kinematic viscosity and the $c_1$, $c_2$, etc., are the corresponding transport coefficients at second order.
Here, we are restricting to flat space\footnote{For curved backgrounds additional terms involving the Riemann tensor appear, see e.g., \cite{Starinets}.}, and we have defined the relativistic shear and vorticity according to
\[
\label{KOdef}
 \K_{ab} = h_a^c h_b^d\D_{(c}u_{d)}, \qquad  \Omega_{ab} = h_a^c h_b^d\D_{[c}u_{d]},
\]
where symmetrisation and anti-symmetrisation are defined as in \eqref{sigmadef}.

Note also we have not included any terms in \eqref{PiT} proportional to some second-order scalar times $h_{ab}$ itself: this is because our definition of the pressure fluctuation $P$ in \eqref{Ft_gauge} forbids all such terms\footnote{Alternative gauge choices for $P$ are however possible.  In particular, one may send $P \rightarrow P - r_c^{3/2}\Pi^\perp/d$, which would render the dissipative part $\Pi^\perp_{ab}$ transverse traceless. This would correspond to the gauge choice \label{gaug_ch} $h^{ab}T_{ab}=d\big(r_c^{-1/2}+r_c^{-3/2}P\big)$.}, as may be seen by going to the rest frame in which $h_{ij}$ reduces to $\delta_{ij}$.

To determine the energy density $\rho$ in the local rest frame, we now simply insert the expansion \eqref{PiT} into the constraint \eqref{rho_result}.
Expanding to second order in fluid gradients, we find
\[
\label{d_result1}
\rho = -\frac{2\eta^2}{p}\K_{ab}\K^{ab}+O(\D^3).
\]

\subsection{Determination of transport coefficients}
\label{Subsection:Hydro_ep}

To test the proposed relativistic hydrodynamic expansion given by \eqref{hydroT}, and to identify the transport coefficients appearing in \eqref{PiT}, we now perform a second expansion of \eqref{PiT} in $\ep$ up to $O(\ep^6$).  We will then backsubstitute into \eqref{hydroT}, and compare with our results for the Brown-York stress tensor in \eqref{T3}, \eqref{T4} and \eqref{T5}.

In comparing the two expansions, it is useful to note that the terms associated with the coefficients $c_5$ and $c_6$ both vanish to $O(\ep^6)$, and so are not constrained by the gravitational analysis above.  Nevertheless, if required, these transport coefficients may be straightforwardly evaluated by extending the gravitational analysis to the appropriate order.

In the following analysis, we expect to be able to recover all of the terms in the Brown-York stress tensor containing up to two gradients, commensurate with the second-order accuracy of our corresponding hydrodynamic expansion.  Note, however, that the number of gradients in a given relativistic term may increase after expanding in $\ep$, since we make free use of the incompressibility and Navier-Stokes equations to simplify our expressions.  Crucially though, the order in gradients can never decrease as we go from the relativistic expansion to the $\ep$-expansion.

Examining the terms up to $\ep^2$ order, as noted earlier we find
\[
\label{P_def}
p = \frac{1}{\sqrt{r_c}} + \frac{P}{r_c^{3/2}},
\]
i.e., the pressure $p$ of the dual fluid consists of an equilibrium part $r_c^{-1/2}$ (as expected from the result for Rindler space) and a fluctuation $r_c^{-3/2}P$ at order $\ep^2$.
The nature of our gauge choice \eqref{Ft_gauge} is such that this relation is exact and receives no corrections at higher orders.
The full hydrodynamical stress tensor \eqref{hydroT} expanded to $\ep^2$ order is then
\begin{align}
T^{\mathrm{hydro}}_{ab}\d x^a \d x^b &= \frac{\d \vec{x}^2}{\sqrt{r_c}}
- \frac{2 v_i}{\sqrt{r_c}}\,\d x^i\d \t
+\frac{v^2}{\sqrt{r_c}}\,\d\t^2 + r_c^{-3/2}\Big[P\delta_{ij}+v_iv_j-2\eta r_c \D_i v_j\Big]\d x^i\d x^j +O(\ep^3),
\end{align}
from which we immediately note $\eta=1$, although we will leave $\eta$ explicit in the following formulae.

Evaluating now the terms at $\ep^3$ order, we find
\[
\label{NewT3}
 T_{ab}^{(3)\,\mathrm{hydro}}\d x^a \d x^b = 2 r_c^{-3/2}\Big[\eta r_c\s_{ik}v_k-(v^2+P)v_i\Big]\d x^i\d\t ,
\]
which is straightforwardly consistent with \eqref{T3} above.

At $\ep^4$ order, we obtain
\begin{align}
T^{(4)\,\mathrm{hydro}}_{ab}\d x^a \d x^b &=  r_c^{-3/2}\Big[v^2(v^2+P)-\eta r_c \s_{ij}v_iv_j - \frac{r_c^2}{2}\sigma_{ij}\sigma_{ij}
\Big]\d\t^2\nn \\[1ex]
&\quad + r_c^{-5/2}\Big[ v_iv_j (v^2+P) +2\eta r_c v_{(i}\p_{j)}P + c_4 r_c^{3/2} \p_i\p_j P+\frac{c_1}{4}r_c^{3/2}\s_{ik}\s_{kj}
\nn\\[1ex]
&\quad\qquad + \frac{c_3}{4}r_c^{3/2} \omega_{ik}\omega_{kj}
-\frac{c_2}{4}r_c^{3/2}\s_{k(i}\omega_{j)k} -2\eta r_c^2 v_{(i}\p^2 v_{j)} -\eta r_c v_{(i}\p_{j)}v^2 \nn\\[1ex]
&\quad\qquad
-\frac{r_c}{2}\eta\s_{ij}v^2  \Big] \d x^i\d x^j \, .\label{newT4}
\end{align}
Comparing with Brown-York stress tensor \eqref{T4}, we find exact agreement upon setting
\[
\label{transport_coefficients}
 c_1= -2\sqrt{r_c}, \quad c_2=c_3=c_4=-4\sqrt{r_c}.
\]
In fact, the only term in \eqref{T4} not captured by the hydrodynamic expansion \eqref{newT4} is the term proportional to $\p^2\s_{ij}$:  as this term is of third order in fluid gradients, however, we do not expect to be able to reproduce it from our second-order hydrodynamic expansion.

Finally, at order $\ep^5$, we obtain
\begin{align}
T^{(5)\,\mathrm{hydro}}_{ab}\d x^a \d x^b &= 2 r_c^{-5/2}\Big[ -c_4r_c^{3/2} v_k \p_k \p_i P + \eta r_c^2 v^2 \p^2 v_i
+\eta r_c^2 v_i v_l \p_k \s_{kl} - \frac{c_2}{8}r_c^{3/2}\s_{ik}\p_k v^2 \nn \\
& \quad +\frac{1}{8}(c_2-2c_1)r_c^{3/2}\s_{ij}\s_{jk}v_k + \frac{c_2}{8}r_c^{3/2}\omega_{ik} \s_{kl}v_l
-\frac{c_3}{4}r_c^{3/2} \omega_{ik}\omega_{kl}v_l \nn\\
&\quad  + \frac{r_c^2}{2}\s_{kl}\s_{kl}v_i
-  v_i v^2 (P+v^2)-\eta r_c v^2\p_i P \nn\\
&\quad
 - \eta r_c v_i v_k \p_k P +\eta r_c \s_{ik}v_k v^2+ \frac{r_c}{2}\eta\omega_{ik}v_k v^2 + \frac{r_c}{2}\eta v_k \s_{kl}v_l v_i  \Big] d\tau dx^i \, .\label{newT5}
\end{align}
Checking this result against \eqref{T5}, we find that we can indeed reproduce all terms (again, apart from the single third-order term proportional to $v_k\p^2\s_{ik}$) with the assignments \eqref{transport_coefficients}.

In conclusion then, we have seen how our seemingly complicated results \eqref{T3}, \eqref{T4} and \eqref{T5} for the Brown-York stress tensor may be recovered from the $\ep$-expansion of the simple relativistic hydrodynamic stress tensor \eqref{hydroT}, where the relevant transport coefficients in \eqref{PiT}  are given by \eqref{transport_coefficients}.

\section{Models for the dual fluid}
\label{Section:Models}

In this section, we present
a simple Lagrangian model for the dual fluid.
Since in general one would not expect to be able to reproduce the dissipative part of the stress tensor from such a model (without first coupling to a heat bath),
we focus on reproducing the non-dissipative piece of the stress tensor,
namely
\[
\label{eqT}
T_{ab} = p h_{ab}.
\]
This corresponds to a fluid with nonzero pressure but a vanishing
energy density in the local rest frame, $\rho = T_{ab} u^{a} u^{b} = 0$.
Note that this stress tensor also satisfies the `equation of state'
following from the Hamiltonian constraint, $d T_{ab} T^{ab} = T^2$.
The equations of motion follow from the conservation
of $T_{ab}$, and are then those in
\eqref{zerograd} but with no higher-derivative corrections, i.e.,
\[ \label{feq}
\partial^a u_a =0, \qquad D u_a =- D^\perp_a \ln p,
\]
where $D^\perp_a \equiv h_a^b\D_b$ and $D \equiv u^a\D_a$.

In the previous section, we showed that the stress
tensor of the dual fluid (including dissipative terms) may be obtained from a
non-relativistic limit of the stress tensor for a relativistic fluid.
The fluid velocity spontaneously breaks relativistic invariance, however,
and so we are led to consider a relativistic Lagrangian in
which the relativistic symmetry is spontaneously broken by a background value
for the field.
The following scalar field action satisfies all requirements,
\[
\label{sqrt_action}
S = \int \d^{d+1}x \sqrt{-\g} \sqrt{-(\D\phi)^2}.
\]
The field equations are given by
\begin{equation}
\nabla^a u_a =0, \qquad  u_a = \frac{\D_a \phi}{\sqrt{X}}, \label{fluid-vec}
\end{equation}
where $X = -(\D\phi)^2$. Note that $u_a$ satisfies,
\[
u^a u_a = -1.
\]
The stress tensor is given by
\[
T_{ab} = \sqrt{X}\g_{ab}+\frac{1}{\sqrt{X}}\D_a\phi\D_b\phi = \sqrt{X} h_{ab},
\]
This is precisely of the form (\ref{eqT}) with $p = \sqrt{X}$
and consequently $T_{ab}$ indeed satisfies the zero energy
density condition, $\rho = T_{ab} u^{a} u^{b} = 0$, and
the quadratic constraint $d T_{ab} T^{ab} = T^2$. By construction, the
stress tensor is conserved and the equations in (\ref{feq}) are
therefore reproduced.

Note that \eqref{fluid-vec} precisely defines the fluid velocity in terms of a potential
(i.e., the fluid motion corresponds to potential flow).
We also note that taking a generic Lagrangian density $\mathcal{L}(X,\phi)$ and then imposing
\[
0 = T_{ab}u^au^b = \(-2\frac{\delta \mathcal{L}}{\delta \g^{ab}}+\g_{ab}\mathcal{L}\)\frac{\D^a\phi\D^b\phi}{X}
= 2X\frac{\delta\mathcal{L}}{\delta X}-\mathcal{L}
\]
uniquely picks out the square root action \eqref{sqrt_action}, up an
integration constant which can be absorbed into a field redefinition.

The equilibrium configuration with pressure $p=r_c^{-1/2}$
in the rest frame\footnote{The background solution corresponding to the equilibrium configuration
in an arbitrary frame is obtained by Lorentz transforming the r.h.s.~of
(\ref{bkd}). Note that in such a case $T_{\tau \tau} \neq 0$ even though the
energy density is zero.} corresponds to the solution,
\[ \label{bkd}
\phi = \t.
\]
(Recall that the background metric is given by
$\g_{ab}\d x^a\d x^b=-r_c\d\t^2+\d x_i\d x^i$.)
This  solution spontaneously breaks Lorentz invariance,
as required.

To model a fluid in its rest frame with small pressure fluctuations about
a constant
equilibrium value of $r_c^{-1/2}$, we must set
\[
\phi = \t + \delta\phi(\t,\vec{x}),
\]
The pressure is then given by
\[
p = \sqrt{X} = \frac{1}{\sqrt{r_c}}(1+2\delta\dot{\phi}+\delta\dot{\phi}^2-r_c\delta\phi_{,i}\delta\phi_{,i})^{1/2} = \frac{1}{\sqrt{r_c}}+\frac{P}{r_c^{3/2}},
\]
where the last equation serves to define the pressure fluctuation $P$ in terms of the field fluctuation $\delta\phi$, in accordance with \eqref{P_def}.
Similarly, from the components of the relativistic fluid velocity $u_\t = -r_c\g=(1+\delta\dot{\phi})/\sqrt{X}$ and
$u_i = \g v_i = \delta\phi_{,i}/\sqrt{X}$, we may solve for the fluid boost velocity $v_i$ in terms of $\delta\phi$, yielding
\[
v_i = -\frac{r_c\delta\phi_{,i}}{(1+\delta\dot{\phi})}.
\]
In summary then, we have shown that the non-dissipative part of the fluid stress tensor \eqref{eqT}, describing a fluid with vanishing
energy density in the local rest frame, leads naturally to the square-root action \eqref{sqrt_action}.
This action is nonlocal in the sense that the expansion around the
background solution involves an infinite number of derivatives.

The square root action has various interesting features.
The action is real for any timelike $(\partial \phi)$ and the Hamiltonian is positive semi-definite since
\[
{\cal H} = X^{-1} (\partial_i \phi)(\partial^i \phi),
\]
so there is no obvious unitarity problem, despite the unconventional nature of the action.
One can generate a wide variety of field theories by adding
extra matter to this scalar field model, provided that one expands about zero background values of these fields.
For example, consider a generic action
\[ \label{gen-mod}
S = \int d^{d+1}x \sqrt{- \gamma} [ f[\psi, A_a, \Phi] \sqrt{- (\partial \phi)^2 + g [\psi,A_a,\Phi]} + h[\psi,A_a,\Phi] ],
\]
with $\psi$ denoting fermions, $A_a$ denoting gauge fields and $\Phi$ denoting scalars. Here the functions
$f[\psi, A_a, \Phi]$, $g [\psi,A_a,\Phi]$ and $h[\psi,A_a,\Phi]$ are only constrained by the requirement that $g$ vanishes
with $f$ finite and $h$ either finite or zero when $\psi = A_a = \Phi = 0$. Choosing a background in which $\phi = \t$ with
all other fields vanishing will always give a zero energy density fluid, as above, for any matter content and choice of functions. Reinstating
the $1/16 \pi G$ in the bulk
will introduce a corresponding prefactor in the action above; as usual, a holographic correspondence
involving classical gravity would be expected to correspond to a dual field theory with a large number of degrees of freedom.

The appearance of the square root in \eqref{sqrt_action} might at first sight suggest a connection with a brane action. Suppose for example that one
considers a $(d+1)$-dimensional brane embedded into a $(d+2)$-dimensional Minkowski target space. The brane embedding would be described by an action
\[
S = - T \int d^{d+1} \xi \sqrt{- \sigma},
\]
where $T$ is the brane tension, $\xi$ are the worldvolume coordinates and $\sigma_{ab} = \D_a Y^{\mu} \D_{b} Y^{\nu} \eta_{\mu \nu}$ is the pulled back
metric, with $Y^{\mu}$ the target space coordinates. Fixing static gauge for the coordinates $(\tau, \vec{x})$ leads to
\[
S = - T \int d^{d+1}x \sqrt{1 + (\D Y)^2 },
\]
where $Y \equiv Y^{d+2}$ is the transverse coordinate to the brane. One can take a tensionless limit of this action in which $T \rightarrow 0$ with
$\varphi = \sqrt{T} Y$ held fixed to eliminate the constant term in the square root
\[
S = - \int d^{d+1}x \sqrt{ (\D \varphi)^2},
\]
but this differs from \eqref{sqrt_action} by the absence of the minus term inside the square root. It follows that the tensionless limit gives rise to a real
action only when $(\D \varphi)^2 > 0$ while the background solution of interest in \eqref{sqrt_action} has $(\D \phi)^2$ timelike, in which case the onshell
brane action would be imaginary. One possibility would be to embed the brane into a flat target space of signature $(d,2)$ so that $Y \rightarrow i Y$
and $\varphi \rightarrow i \varphi \equiv \phi$, reproducing \eqref{sqrt_action} in the tensionless limit.

\

\section{Discussion}
\label{Discussion}

In this paper we established a direct relation between $(d+2)$-dimensional
Ricci-flat metrics and $(d+1)$-dimensional fluids satisfying
the incompressible Navier-Stokes equations, corrected by specific higher-derivative terms.
Our results raise many interesting questions and there are
diverse directions one may wish to further pursue.

Perhaps the most interesting question is whether the correspondence
extends beyond the hydrodynamic regime (on the field theory side)
and/or the classical gravitational description (on the bulk side).
Is there a string embedding of this correspondence?
Even without a string embedding, one may ask how this
correspondence changes if one adds, for example, a bulk stress
tensor or considers higher-derivative corrections to
Einstein gravity.
Will such changes modify the properties of the dual fluid?

The dual fluid has many unconventional features. In particular,
it has zero energy density in equilibrium but nonzero pressure.
What theories have such properties?
One of our most tantalising observations is that there exists
a simple scalar field model that has these properties
and at the same time has no obvious problems with
unitarity, etc. This provides then a candidate model for
the dual to flat spacetime.  Can one obtain this model
from branes? Are there any other theories which at and near
equilibrium are described by such a fluid?

This paper has focused on the gravity/fluid
correspondence, computing transport coefficients of the dual fluid
holographically, but it is very exciting to ask how far flat space
holography can be developed. Defining the holographic theory on
$\Sigma_c$, is there a holographic dictionary relating bulk
computations to quantities computable in the dual field theory model?

Here, the construction of the bulk metric was achieved using a
non-relativistic hydrodynamic expansion. We have seen, however,
that there is an underlying relativistic description. Can one
find a manifestly relativistic construction of the bulk metric?

The metric perturbations are by construction regular at the horizon and
throughout the region $r_h < r < r_c$, but, in general, one would expect that the
series expansion only converges in a finite neighbourhood
of the surface $\Sigma_c$. What is the dependence of the radius of convergence
on the defining data on $\Sigma_c$, and what is the range of validity
of the gauge choice for the radial foliation off the surface $\Sigma_c$?
Can one resum the series to obtain a closed-form expression for the
metric?

It is important to note
here that the hydrodynamic expansion can be made around a surface of
{\it arbitrary} radius $r_c$, which in particular can be far from the
horizon. (As $r_{c} \rightarrow \infty$, the pressure and temperature
of the equilibrium fluid tends to zero; the stress tensor of the fluid
is then zero in this limit).  The expansion is therefore in general
physically distinct from a near-horizon expansion, although it would
certainly be useful to understand in more detail the limit in which
$r_c$ is taken to be close to the horizon, following \cite{Andy}.

In this paper, we have used flat space in Rindler coordinates as a
seed solution about which the hydrodynamic expansion is made.
 As the present construction uses neither the existence of
an event horizon, nor the detailed asymptotic structure of flat space,
there is no reason why it should be applicable only to asymptotically
flat or black hole spacetimes.
In fact, by the equivalence principle,
the construction should hold locally in any small neighbourhood.
Can one patch such a `local' holographic description of neighbourhoods
to obtain a global holographic description of general
spacetimes?

Another generalisation would be to start from a different seed solution.
This should admit at least one Killing vector for the solution to support
equilibrium configurations, but the dual fluid need not live on a flat
spacetime; it could live instead, for example,
on a background such as $R_{\tau} \times S^{d}$. Note that this
would be the geometry of constant-radius slices of Schwarzschild
spacetime.

This work also raises many interesting questions from a general
relativity viewpoint. We have shown explicitly that the data defined
on $\Sigma_c$, together with regularity at the horizon, suffices to
define the solution uniquely as a series expansion in the region
between $\Sigma_c$ and the horizon. Do these solutions exist globally?\footnote{Note that the same question applies also to the standard AdS/fluid
correspondence. In exact parallel, it was shown
in \cite{Bhattacharyya:2008jc}
 that the derivative expansion defines the solution uniquely as
a series expansion in the region between the conformal boundary and the
future horizon. It is a non-trivial question whether this provides a global
solution. We thank Bob Wald for discussions regarding this point.}
A priori it would seem
far from obvious that this radial initial value problem is well-posed
and admits a unique solution: it would be instructive to understand
better these issues.

All in all, this correspondence provides an interesting arena
to explore holography for general spacetimes and raises many
interesting questions that we hope to return to in the future.

\acknowledgments

The authors wish to thank Jan de Boer and Michal Heller for discussions.
This work is part of the research program of the `Stichting voor
Fundamenteel Onderzoek der Materie' (FOM), which is financially
supported by the `Nederlandse Organisatie voor Wetenschappelijk
Onderzoek' (NWO). The authors acknowledge support from NWO;
GC and KS via an NWO Vici grant, and PM via an NWO Veni grant.

\appendix

\section{Choice of radial gauge}
\label{app:radialgauge}

The boundary conditions imposed on the surface $\Sigma_c$ defined by $r=r_c$ do not constrain the radial coordinate away from the surface $\Sigma_c$. In the main text, we adopt coordinates such that lines of constant $x^a$ form a null geodesic congruence around $\Sigma_c$. Moreover, along each null geodesic, we will choose the transverse coordinate $r$ to be an affine parameter.

These geometrical conditions can be translated into conditions on the metric components as follows.
Firstly, the tangent to lines of constant $x^a$ is $n = f(x^a,r)\partial_r$. Here, $n$ is a null vector if and only if $g_{rr} = 0$. One then has
\[
n^\nu \nabla_\nu n^\mu = n^\mu \p_r f  + f^2 g^{\mu a}\p_r g_{ra}.
\]
Hence, vectors $n$ are tangent to geodesics with affine parameter $r$ if and only if $\p_r g_{ra} = 0$ and $f(x^a,r) = f(x^a)$. In summary, we impose the conditions
\[
g_{r r} = 0,\qquad \partial_r g_{r a} = 0.\label{gauge1}
\]
These conditions are similar to those used in the conformal fluid/anti-de Sitter gravity correspondence (see e.g., \cite{Bhattacharyya:2008jc,Bhattacharyya:2008ji}). Note that there is in general no guarantee that these coordinates may be extended far away from $\Sigma_c$. Rather, these coordinates might break down at some finite radius outside $\Sigma_c$ if the null congruence has caustics.

\section{Infinitesimal diffeomorphisms preserving equilibrium}
\label{app:no_rho}

Let us consider a general infinitesimal diffeomorphism around Rindler spacetime. From the boundary condition $g_{ab} = \g_{ab}$ at $r=r_c$, we find that on the surface $r=r_c$ the following conditions hold
\begin{align}
\label{zeq1}
\p_{(i} \xi_{j)} &= 0, \\
\label{zeq2}
\p_i \xi^r - r_c \p_i \xi^\tau + \p_\tau \xi^i &= 0,  \\
\label{zeq3}
-\xi^r + 2\p_\tau (\xi^r - r_c \xi^\tau)&= 0.
\end{align}
The first of these equations amounts to Killing's equation on $d$-dimensional Euclidean space and is solved by
\[
\xi_{i} = a_i(r,\tau) + b_{[ij]}(r,\tau) x^j.
\]
In particular, we obtain $\partial_i \xi_i = 0$ and $\partial_j \partial_j \xi_i = 0$. Acting on \eqref{zeq3} with $\p_i$ and using \eqref{zeq2}, we obtain
\[
\p_i \xi^r = -2 \p_\tau^2 \xi^i.
\]
These equations are integrable for $\xi^r$ only if $\p_\tau^2 b_{[ij]} = 0$. We then find
\bea
b_{[ij]}(r,\tau) &=& \tau d_{[ij]}(r) + e_{[ij]}(r) \\
\xi^r &=& -2 \p_\tau^2 a_i(r,\tau) x^i + c(r,\tau )\, .
\eea
The equation \eqref{zeq2} is integrable only if $d_{[ij]} =0$. The equations \eqref{zeq2}-\eqref{zeq3} are then solved by
\[
\xi^\tau = \big(-\frac{2}{r_c}\p_\tau^2 a_i + \frac{1}{r_c}\partial_\tau a_i\big)x^i + \frac{c(r,\tau)}{r_c}-\frac{1}{2r_c}\int d\tau c(r,\tau).
\]
Finally, the diffeomorphisms preserving the induced metric on $\Sigma_c$ are given by
\bea
\xi^r &=& -2 \p_\tau^2 a_i(r,\tau) x^i + c(r,\tau ) ,\nn \\
\xi^\tau &=& \big(- \frac{2}{r_c}\p_\tau^2 a_i (r,\tau)+ \frac{1}{r_c}\partial_\tau a_i(r,\tau)\big)x^i + \frac{c(r,\tau)}{r_c}-\frac{1}{2r_c}\int d\tau c(r,\tau)+t(r) , \nn\\
\xi_{i} &=& a_i(r,\tau) + e_{[ij]}(r) x^j.
\eea
One then computes the Brown-York stress tensor on $\Sigma_c$ for the metric $g_{\mu\nu} = \bar g_{\mu\nu}+\mathcal L_\xi \bar g_{\mu\nu}$.
The resulting stress tensor has the form of a relativistic perfect fluid $T_{ab} = \rho u_a u_b + p (\g_{ab} + u_a u_b)$, where
\bea
\rho &=& 0 \nn\\
p &=& \frac{1}{\sqrt{r_c}} + \frac{1}{2r_c^{3/2}}(-c +4\p_\tau^{2}c -2 x^i \partial_\tau v_i) |_{r=r_c} \nn\\
v_i &=& - \p_\tau a_i +4 \p_\tau^{3} a_i |_{r=r_c} \, .
\eea
We have therefore shown that a nonzero energy density cannot be generated from an infinitesimal diffeomorphism acting on Rindler spacetime. Note that we
did not need to use the gauge condition \eqref{gauge1} for this argument.

Let us now require that the fluid is in equilibrium with a uniform velocity profile $v_i$ and pressure $p$ on $\Sigma_c$. For convenience, let us define $r_h$ via $p=r_c^{-1/2}(1+r_h/2r_c)$. The set of allowed diffeomorphisms then reduces to
\bea
a_i(r,\tau) &=& A_i(r)-v_i(r)\tau +e^{\tau/2}B_i(r)+e^{-\tau/2}C_i(r) ,\nn\\
c(r,\tau)&=& -R(r)+e^{\tau/2}D(r)+e^{-\tau/2}E(r),
\eea
where $v_i(r_c) = v_i$ and $R(r_c) = r_h$. We now fix the gauge freedom associated with redefining the surfaces of constant $r$ by imposing the conditions \eqref{gauge1}.  Stationarity and homogeneity then further restrict $\partial_b g_{ra}(r,x^c) = 0$. The vectors obeying these conditions reduce simply to a linear combination of a finite set of transformations,
\bea
\xi^r &=& -r_h +e^{\tau/2}\xi_{(0)}^v - \frac{1}{2}e^{\tau /2}B^i x_i \,, \nn\\
\xi^\tau &=& \xi^\tau_{(0)} + \frac{r_h}{2r_c}\tau - \frac{v_i x^i}{r_c} \,, \nn\\
\xi^i &=& \xi_{(0)}^i + \omega_{[ij]}x^j - v_i \tau +f^i r + e^{\tau/2}B^i\, .
\eea
Recall that Rindler spacetime can be rewritten in null coordinates $u=e^{\tau/2}$, $v=4r e^{-\tau/2}$. One then recognises $\tau$-translations or alternatively boosts in the $(u,v)$ plane (associated with $\xi^\tau_{(0)}$), spatial translations (associated with $\xi_{(0)}^i$), rotations (associated with $\omega_{[ij]}$), $v$-translations  (associated with $\xi_{(0)}^v$) and boosts (associated with $B^i$), all of which are Killing symmetries and do not lead to any perturbation $\mathcal L_\xi \bar g_{\mu\nu}$. The $r$-dependent translation (associated with $f^i$) amounts solely to shifting $g_{r i}$ by $f^i$.
If we restrict the gauge choice further so that $g_{ri} = -v_i/r_c$  (as in \eqref{gr_components}), this then fixes $f^i=0$.

Finally, the nontrivial perturbations are (i) a shift of $r$ combined with a linearised dilatation of time (associated with $r_h$); (ii) a linearised boost in the $i$ direction (associated with $v_i$). These two transformations are the linearisation of the two finite transformations \eqref{t1}-\eqref{t2} considered in the main text. The pressure $p=r_c^{-1/2}(1+r_h/2r_c)$ is the linearisation of the pressure \eqref{pressequil} in the main text.

\section{A basis for second-order relativistic hydrodynamics}
\label{hydro_app}

In this appendix we show how to construct a general basis for the scalars and transverse tensors that appear at second order in the gradient expansion
of relativistic hydrodynamics, for the special case in which the equilibrium energy density vanishes.  We restrict to the case in
which the dual fluid is in a flat background.

We begin by writing down all possible symmetric transverse tensors constructed from two gradients of the fluid pressure $p$ and velocity $u^a$.  These
take the form $X^\perp_{ab} = h_{(a}^c h_{b)}^d X_{cd}$ where the transverse projector $h^a_b=\delta^a_b+u^au_b$, and the possible independent choices for $X_{cd}$ are
\begin{align}
\label{Xlist}
& \D_c\ln p\,\D_d\ln p, \qquad \D_c\D_d \ln p, \qquad u^e\D_e \ln p\, \D_c u_d, \qquad u^e\D_e u_c\D_d\ln p, \nn\\
& u^e\D_e\D_c u_d, \qquad \D_e u^e \D_c u_d, \qquad \D_e u_c\D^eu_d, \qquad \D_e u_c\D_du^e, \qquad \D_c u_e\D_d u^e
\end{align}
(Note that the term $u^e\D_c\D_d u_e = -\D_c u^e\D_d u_e$ and so is not independent.)
We have not considered terms consisting of the induced metric $h_{ab}$ multiplied by a scalar function of second order in gradients since our choice of gauge \eqref{Ft_gauge} effectively absorbs all such corrections into the zeroth order $p h_{ab}$ term in \eqref{hydroT}.

We may now reduce this list to just six independent choices by applying the fluid equations of motion evaluated to leading order in gradients, given in equation \eqref{zerograd}.  In particular, these tell us that the fluid divergence $\K=\D_au^a$ vanishes at first order in the gradient expansion, meaning the term $\D_e u^e \D_c u_d$ in the list above vanishes at second order and may be dropped.
Using the identity
\[
\label{Du_id}
 \D_a u_b = \K_{ab}+\Omega_{ab}-u_aDu_b,
\]
where the transverse tensors $\K_{ab}$ and $\Omega_{ab}$ are defined in \eqref{KOdef} and the fluid derivatives $D \equiv u^a\D_a$ and $D^\perp_a\equiv h_a^b\D_b$,
the six remaining independent choices for $X^\perp_{ab}$ may be written as
\[
\label{tensor_list}
\K_a^c\K_{cb}, \qquad  \K_{(a}^c\Omega_{|c|b)}, \qquad \Omega_a^{\,\,\,c}\Omega_{cb}, \qquad
h_a^ch_b^d\D_c\D_d \ln p, \qquad D^\perp_a \ln p \,D^\perp_b\ln p, \qquad  \K_{ab}\,D\ln p .
\]
For example, the term $h_{(a}^ch_{b)}^d u^e\D_e\D_cu_d$ reduces to a linear combination of the last three terms in the list above, by reversing the order of the derivatives, then pushing the $u^e$ inside the $\D_c$ and applying \eqref{zerograd}.

Although in the present case the energy density $\rho$ in the local rest frame is fixed by the constraint \eqref{rho_result}, as a matter of general interest one might wish to construct a complete basis for second-order scalar quantities, in a fashion analogous to that done above for symmetric transverse tensors.

One way to construct such a basis is to contract the tensors appearing in \eqref{Xlist}, and then apply the leading order equations of motion \eqref{zerograd} along with the identity \eqref{Du_id}.  After eliminating all dependencies, we find the following complete set of independent terms:
\[
\label{scalar_list}
\K_{ab}\K^{ab}, \qquad \Omega_{ab}\Omega^{ab}, \qquad
 D^\perp_a\ln p \,D^{\perp a}\ln p,  \qquad D\ln p \,D\ln p, \qquad D^2 \ln p.
\]
Note that taking the divergence of the leading order equations of motion \eqref{zerograd} yields the result
\[
 h^{ab}\D_a\D_b\ln p = D^{\perp a}\ln p D^\perp_a \ln p -K_{ab}K^{ab} + \Omega_{ab}\Omega^{ab} + O(\D^3),
\]
which, upon expanding in $\ep$, reduces to our earlier result \eqref{derived_eqn}.

Using this list of terms, for a general hydrodynamic theory with vanishing equilibrium energy density, one could then write
\[
\label{hydrorho}
 \rho = b_1 \K_{ab}\K^{ab}+b_2 \Omega_{ab}\Omega^{ab}+ b_3 D\ln p \,D\ln p + b_4 D^2 \ln p+b_5 D^\perp_a\ln p \,D^{\perp a}\ln p  + O(\D^3),
\]
where $b_1$, $b_2$, etc., define additional second-order transport coefficients.


\begin{thebibliography}{99}

\bibitem{BS} R. Beig and B. Schmidt, {\it Einstein's equations near spatial
infinity, Commun.\ Math.\ Phys.}\ {\bf 87} (1982) 65.

\bibitem{BS2} R. Beig,
{\it Integration of Einsteins Equations Near Spatial Infinity,
Proc.\ R.\ Soc.\ A} {\bf 391} (1984) 295.

\bibitem{deHaro:2000wj}
  S.~de Haro, K.~Skenderis, S.~N.~Solodukhin,
  {\it Gravity in warped compactifications and the holographic stress tensor,
  Class.\ Quant.\ Grav.}\  {\bf 18} 3171-3180,
  [hep-th/0011230].

\bibitem{Skenderis:2002wp}
  K.~Skenderis,
  {\it Lecture notes on holographic renormalisation,
  Class.\ Quant.\ Grav.}\  {\bf 19 } (2002)  5849-5876,
  [hep-th/0209067].

\bibitem{Papadimitriou:2010as}
  I.~Papadimitriou,
  {\it Holographic renormalisation as a canonical transformation,}
  [arXiv:1007.4592].

\bibitem{Bredberg}
  I.~Bredberg, C.~Keeler, V.~Lysov and A.~Strominger,
  {\it From Navier-Stokes To Einstein,}
  [arXiv:1101.2451].

\bibitem{Andy}
  I.~Bredberg, C.~Keeler, V.~Lysov and A.~Strominger,
  {\it Wilsonian Approach to Fluid/Gravity Duality,}
  [arXiv:1006.1902].

\bibitem{Fouxon:2008tb}
  I.~Fouxon and Y.~Oz,
  {\it Conformal Field Theory as Microscopic Dynamics of Incompressible Euler and
  Navier-Stokes Equations,
  Phys.\ Rev.\ Lett.}\  {\bf 101} (2008) 261602,
  [arXiv:0809.4512].

\bibitem{Bhattacharyya:2008kq}
  S.~Bhattacharyya, S.~Minwalla and S.~R.~Wadia,
  {\it The Incompressible Non-Relativistic Navier-Stokes Equation from Gravity,
  JHEP} {\bf 0908} (2009) 059,
  [arXiv:0810.1545].

\bibitem{Eling:2009pb}
  C.~Eling, I.~Fouxon and Y.~Oz,
  {\it The Incompressible Navier-Stokes Equations From Membrane Dynamics,
  Phys.\ Lett.\  B} {\bf 680} (2009) 496,
  [arXiv:0905.3638].

\bibitem{Padmanabhan}
 T.~Padmanabhan,
{\it Entropy density of spacetime and the Navier-Stokes fluid
dynamics of null surfaces,  Phys.~Rev.~D} {\bf 83} (2011) 044048,
[arXiv:1012.0119].

\bibitem{Damour1}
T.~Damour,
{\it Quelques propri{\'e}t{\'e}s m{\'e}caniques, {\'e}lectromagn{\'e}tiques, thermodynamiques et
quantiques des trous noirs,} Th{\`e}se de Doctorat d'Etat, Universit{\'e} Pierre et Marie Curie, Paris VI,
1979.

\bibitem{Damour2}
T.~Damour,
{\it Surface effects in black hole physics,
Proceedings of the second Marcel Grossmann meeting on General Relativity} (1982),
Ed.~R.~Ruffini, North Holland.

\bibitem{Thorne}
K.~S.~Thorne, R.~H.~Price, and D.~A.~Macdonald,
{\it Black Holes: the Membrane Paradigm}, Yale University Press,  New
Haven, USA (1986).

\bibitem{Bhattacharyya:2008jc}
  S.~Bhattacharyya, V.~E.~Hubeny, S.~Minwalla and M.~Rangamani,
  {\it Nonlinear Fluid Dynamics from Gravity,
  JHEP} {\bf 0802} (2008) 045,
  [arXiv:0712.2456].

\bibitem{Landau} L.~Landau and E.~Lifshitz, {\it Fluid Mechanics: Course of Theoretical Physics Vol.~6}, Elsevier Butterworth-Heinemann (1959), pg.~513.

\bibitem{Niayesh1}
N.~Afshordi, D.~J.~H.~Chung, G.~Geshnizjani,
{\it Cuscuton: a causal field theory with an infinite speed of sound,
Phys.~Rev.~D} {\bf 75} (2007) 083513,
[hep-th/0609150].

\bibitem{Niayesh2}
N.~Afshordi, D.~J.~H.~Chung, M.~Doran and G.~Geshnizjani,
{\it Cuscuton cosmology: dark energy meets modified gravity,
Phys.~Rev.~D} {\bf 75} (2007) 123509,
[astro-ph/0702002].

\bibitem{Brown:1992br}
  J.~D.~Brown and J.~W.~York,
  {\it Quasilocal energy and conserved charges derived from the gravitational
  action,
  Phys.\ Rev.\  D} {\bf 47} (1993) 1407,
  [gr-qc/9209012].

\bibitem{Bardeen:1973gs}
  J.~M.~Bardeen, B.~Carter, S.~W.~Hawking,
  {\it The Four laws of black hole mechanics,
  Commun.\ Math.\ Phys.}\  {\bf 31} (1973)  161-170.

\bibitem{Townsend:1997ku}
  P.~K.~Townsend,
  {\it Black holes: Lecture notes,}
    [gr-qc/9707012].

\bibitem{Bhattacharyya:2008xc}
  S.~Bhattacharyya {\it et al.},
  ``Local Fluid Dynamical Entropy from Gravity,''
  JHEP {\bf 0806} (2008) 055
  [arXiv:0803.2526 [hep-th]].

\bibitem{Starinets}
  R.~Baier, P.~Romatschke, D.~T.~Son, A.~O.~Starinets and M.~A.~Stephanov,
  {\it Relativistic viscous hydrodynamics, conformal invariance, and holography,
  JHEP} {\bf 0804} (2008) 100,
  [arXiv:0712.2451].

\bibitem{Bhattacharyya:2008ji}
  S.~Bhattacharyya, R.~Loganayagam, S.~Minwalla, S.~Nampuri, S.~P.~Trivedi, S.~R.~Wadia,
  {\it Forced Fluid Dynamics from Gravity,
  JHEP} {\bf 0902} (2009)  018,
  [arXiv:0806.0006].


\end{thebibliography}
\end{document}